\documentclass[]{aa}  

\usepackage{graphicx}
\usepackage{txfonts}
\usepackage{mathtools}
\usepackage{amsmath}
\usepackage{pdfpages}
\usepackage{natbib}
\usepackage{url}
\usepackage{placeins}
\usepackage{tikz}
\usepackage{wasysym}
\usepackage{ulem}
\usetikzlibrary{shapes,arrows}

\newcommand{\eq}{eq. }

\newcommand{\sect}{Section }
\newcommand{\fig}{Figure}

\newcommand{\Bv}{\textbf{B}}

\newcommand{\Bm}{\overline{B}}
\newcommand{\Bmx}{\overline{B}_x}
\newcommand{\Bmy}{\overline{B}_y}

\newcommand{\Jmx}{\overline{J}_x}
\newcommand{\Jmy}{\overline{J}_y}

\newcommand{\Hm}[1] {\overline{#1}}
\newcommand{\Vm}[1] {\langle{#1}\rangle}
  
  \newcommand{\Vmt}[1] {\langle\langle{#1}\rangle\rangle_t}
\begin{document}

   \title{Magnetorotational instability with smoothed particle hydrodynamics}

   \author{Robert Wissing, Sijing Shen, James Wadsley and Thomas Quinn
          }
   \institute{
   Institute of Theoretical Astrophysics, University of Oslo, Postboks 1029, 0315 Oslo, Norway \\ \email{robertwi@astro.uio.no; sijing.shen@astro.uio.no; wadsley@mcmaster.ca; trq@astro.washington.edu}}
   \date{}
 
  \abstract
   { 
   The magnetorotational instability(MRI) is an important process in driving turbulence in sufficiently-ionized accretion disks. It has been extensively studied using simulations with Eulerian grid codes, but remains fairly unexplored for meshless codes. Here, we present a thorough numerical study on the MRI using the smoothed particle magnetohydrodynamics method (SPMHD) with the geometric density average force expression (GDSPH). We perform a plethora of shearing box simulations with different initial setups and a wide range of resolution and dissipation parameters. We show, for the first time, that MRI with sustained turbulence can be simulated successfully with SPH, with results consistent with prior work with grid-based codes, including saturation properties such as magnetic and kinetic energies and their respective stresses. In particular, for the stratified boxes, our simulations reproduce the characteristic "butterfly" diagram of the MRI dynamo with saturated turbulence for at least 100 orbits. On the contrary, traditional SPH simulations suffer from runaway growth and develop unphysically large azimuthal fields, similar to the results from a recent study with mesh-less methods. We investigated the dependency of MRI turbulence on the numerical Prandtl number($P_m$) in SPH, focusing on the unstratified, zero net-flux case. We found that turbulence can only be sustained with a Prandtl number larger than $\sim 2.5$, similar to the critical values of physical Prandtl number found in grid-code simulations. However, unlike grid-based codes, the numerical Prandtl number in SPH increases with resolution, and for a fixed Prandtl number, the resulting magnetic energy and stresses are independent of resolution. Mean-field analyses were performed on all simulations, and the resulting transport coefficients indicate no $\alpha$-effect in the unstratified cases, but an active $\alpha \omega$ dynamo and a diamagnetic pumping effect in the stratified medium, which are generally in agreement with previous studies. There is no clear indication of a shear-current dynamo in our simulation, which is likely to be responsible for a weaker mean-field growth in the tall, unstratified, zero net-flux simulation. 
  }


   \maketitle
%

\section{Introduction}
A popular mechanism for the generation of turbulence within accretion disks is the magnetorotational instability (MRI; \cite{velikhov1959stability,1960PNAS...46..253C,1969A&A.....1..388F,1991ApJ...376..214B}), which is a local linear instability that occurs for magnetic fields in Keplerian-like flows (e.g. accretion disks). The ensuing turbulence subsequently acts as a driver for angular momentum transport within the disk, allowing efficient mass accretion onto the central object \citep{1973A&A....24..337S,1974MNRAS.168..603L}. Due to its simple prerequisites of "activating" the instability (negative angular momentum gradient, weak magnetic field, and sufficiently ionized gas), the MRI is potentially a crucial component in many different astrophysical systems.  
\\ \\
While the linear behavior of the MRI is well established \citep{1991ApJ...376..214B,1992ApJ...400..610B,1994ApJ...434..206C,1994ApJ...432..213G,2004ApJ...602..892K}, the non-linear phase is an active area of research. Modeling the full non-linear behavior of the MRI requires numerical simulations, which for the past few decades have been readily applied to study the MRI in both global \citep{2010MNRAS.408..752P,2011ApJ...738...84H,2013ApJ...772..102H,2013MNRAS.435.2281P,2020ApJ...891..154D} and local \citep{1995ApJ...440..742H} setups. While global simulations allow the inclusion of global properties such as winds, jets, and accretion, they require extensive computational resources to properly resolve the MRI growth rates. \cite{2004ApJ...605..321S} found that a minimum of six grid zones per MRI wavelength ($\lambda_{MRI}$) was required to model the linear phase. This criterion was further extended to the non-linear regime by \cite{2010ApJ...711..959N}, where an effective quality parameter ($Q$) was used to gauge the resolution requirement for correct MRI behavior. On the other hand, local setups allow higher resolution and remain simple and well-posed for investigating the non-linearity and saturation of the MRI under specific initial conditions. In general, local setups apply a shearing box approximation \citep{1965MNRAS.130..125G,1995ApJ...440..742H}, which models either a small unstratified block of gas within the disk (neglecting gravity) or a slice of the disk with a vertically stratified density field (including the vertical gravity component). The initial configuration of the magnetic field is an important factor for the behavior and saturation of the MRI, and most studies apply either a constant mean magnetic field through the box (often referred to as the net flux case [NF; \cite{1995ApJ...440..742H,2004ApJ...605..321S,2009ApJ...694.1010G,2009ApJ...690..974S}]) or an initial magnetic field which has a zero mean-field value (often referred to as the zero net flux case [ZNF;\cite{1996ApJ...464..690H,2007A&A...476.1113F,2009ApJ...707..833S,2011ApJ...739...82B}]. Both cases are idealized setups. In reality, the mean-field within local patches of the disk (with sizes around the scale height of disk) will vary in time due to larger-scale current structures and non-ideal MHD effects. In general, the initial mean-field within local setups is either along azimuthal or vertical direction, as any radial mean-field will lead to a constant increase in the azimuthal mean-field due to the shear.
\\ \\
Within less ionized accretion disks such as protoplanetary disks, non-ideal effects become important. Simulations including non-ideal MHD, indicate that the MRI within protoplanetary disks struggles to generate enough angular momentum transfer on its own to fit observed values \citep{1996ApJ...457..355G,2003ApJ...585..908F,2008ApJ...679L.131T,2011ApJ...732L..30H,2013ApJ...764...66S,2013ApJ...772...96B,2014A&A...566A..56L,2014ApJ...791..137B,2015ApJ...798...84B}. However, global effects on the MRI such as magnetic winds \citep{2013ApJ...769...76B,2015MNRAS.454.1117S,2015ApJ...801...84G,2017A&A...600A..75B} and gravitational instability \citep{2020ApJ...891..154D} can significantly alter the effective angular momentum transport within disks.
\\ \\
The turbulence generated by the MRI is sub-critical, which means that it requires a self-sustaining process to remain active \citep{2007PhRvL..98y4502R,2011PhRvE..84c6321H,2013JFM...731....1R}. A physical explanation for the turbulence saturation in the vertical net-flux case was proposed by \citet{1994ApJ...432..213G}, \citet{2009ApJ...698L..72P} and \citet{2009MNRAS.394..715L}, in which saturation is driven by parasitic (secondary) instabilities that break down the so-called channel modes (axisymmetric radial streaming motions) generated by the primary (fastest-growing) MRI modes. These correspond to both Kelvin-Helmholtz instabilities which feed off the shear in the velocity field and tearing mode instabilities which feed off the current density. The secondary instabilities themselves eventually decay into small-scale turbulence which then, in combination with the vertical net-flux, regenerate the MRI modes, creating a self-sustaining loop. The mechanism for saturation becomes more difficult to pinpoint when there is no global mean-field in the box (ZNF), since both the magnetic field and turbulence are required to sustain each other. In addition, the unstratified ZNF case is statistically symmetric, which means that there are no net helicities within the flow, and this makes the generation of local mean-fields more difficult. However, dynamo cycles and coherent local mean-field growth have been observed in previous simulations of the unstratified ZNF case \citep{2016MNRAS.456.2273S}. The underlying process of growth still remains uncertain, but the two primary theories are the stochastic alpha effect \citep{1997ApJ...475..263V,2000A&A...364..339S,2011PhRvL.107y5004H} and the magnetic shear current effect \citep{2003PhRvE..68c6301R,2004PhRvE..70d6310R,2015PhRvL.114h5002S}, which we will examine in more detail in the upcoming sections.
\\ \\
Adding stratification to the shearing box represents a more realistic view of the accretions disks and brings forth many new mechanisms that act on the behavior and saturation of the system. The stratified case enables buoyancy instabilities, which transports magnetic fields from the outer central region upwards. Beyond $|z|>2H$, the gas is magnetically dominated, less turbulent and buoyantly unstable (e.g., \cite{2010ApJ...708.1716S,2011ApJ...728..130G}). Meanwhile, the sign of the field flips within the mid-plane and this becomes a cyclical behavior that occurs around every 10 orbits (producing the characteristic butterfly diagram). This behavior of the magnetic field indicates an active mean-field dynamo (e.g., \cite{1995ApJ...446..741B,1996ApJ...463..656S,2006ApJ...640..901H,2010MNRAS.405...41G,2010ApJ...713...52D,2011ApJ...730...94S}). The stratified disk dynamo is complicated, likely involving several mechanisms acting together; as such the cyclic behavior observed within these simulations still remains unclear. While the buoyancy instabilities dominate the regions beyond the scale height, the central region remain buoyantly stable, requiring an alternative mechanism in this region \citep{2010ApJ...708.1716S,2010MNRAS.405...41G}. One such mechanism is through the alpha-effect, or more explicitly the outward transport of small scale magnetic helicity flux\citep{2001ApJ...550..752V,2004PhRvL..93t5001S}. Another mechanism that can advect magnetic fields is turbulent pumping which can expel magnetic field from high turbulent regions to lower turbulent regions \citep{2010MNRAS.405...41G}. In addition, the dynamo mechanisms in the unstratified case likely plays a role here too, as the unstratified case represents an approximation of the mid-plane in the disk \citep{2008A&A...488..451L,2011MNRAS.413..901K}.
\\ \\
The stratified shearing box is also dependent on the strength and geometry of the global magnetic mean-field, showing a wide range of different behaviors. For example, compared to an azimuthal mean-field, the presence of a vertical mean-field greatly enhances the stress within the fluid and exhibits powerful outflows which can increase the removal of angular momentum from the disk \citep{2009ApJ...691L..49S,2011ApJ...728..130G,2011ApJ...730...94S,2011ApJ...736..144B,2013ApJ...764...66S}.
\\ \\ 
Seminal work by \cite{2007A&A...476.1123F} showed that for the ZNF unstratified case the saturated turbulence level decreased with higher resolution. This highlighted the importance of small-scale dissipation for the MRI. Further work has shown that more or less all MRI cases are sensitive to the small-scale dissipation, where kinematic viscosity ($\nu$) and magnetic resistivity ($\eta$) play a major role. The ratio between the two, the so-called magnetic Prandtl number ($P_m=\nu/\eta$), is shown to be fundamentally important in determining the MRI saturation and the stress, and in general the behavior of MHD turbulence in any system \citep{2004ApJ...612..276S,2004PhRvL..92e4502S,2014ApJ...797L..19F,2016JPlPh..82f5301F}. In nature, galaxies, galaxy clusters and molecular clouds have magnetic Prandtl numbers far greater than unity ($P_m >> 1$). For example, in molecular clouds $P_m\approx10^{10}$ \citep{2016JPlPh..82f5301F}. On the opposite extreme, protostellar discs and stars usually have magnetic Prandtl numbers much smaller than unity ($P_m << 1$) \citep{2005PhR...417....1B,2007NJPh....9..300S}. In MRI simulations, the Prandtl number can be either physical, where explicit dissipation is added to the system, or numerical, which is determined by the numerical dissipation of the numerical scheme. Higher Prandtl numbers generally increase the angular momentum transport \citep{2007A&A...476.1113F,2007MNRAS.378.1471L,2009ApJ...690..974S,2009ApJ...707..833S,2010EAS....41..167F}. In the low Prandtl number limit, while the NF case still exhibits MRI turbulence and saturates at a low but finite value of angular momentum transport \citep{2015A&A...579A.117M}, in the ZNF case turbulence cannot be sustained for Prandtl numbers below a certain critical value ($P_m<2$ in \cite{2007A&A...476.1123F})\footnote{Note that the critical Prandtl number is dependent on the Reynolds number, however no study have found a critical Prandtl number lower than $P_{m,c}=2$ for the standard box.}. In addition, the convergence behavior of the MRI turbulence and the critical Prandtl number can also be sensitive to the vertical aspect ratio of the domain: while simulations with standard box (with vertical-over-radial aspect ratio, $L_z/L_x = 1$) exhibit decreased stress levels with increasing resolution, in the tall-box simulations ($L_z/L_x>2.5$) the stress levels are converged. The stress saturation still depends on the Prandtl number in the tall boxes, albeit with a somewhat lower critical value and with longer lifetimes \citep{2016MNRAS.456.2273S}. 
\\ \\ 
While a lot of the focus surrounding the Prandtl number has revolved around the physical Prandtl number, not many studies have been done on the numerical Prandtl number. Numerical dissipation acts differently compared to physical dissipation, depending heavily on the fluid flow and the resolution. How well the numerical Prandtl number relates to the observed dependency on the physical Prandtl number for MHD turbulence is still unclear, but a similar dependency is expected. The consequence of not knowing the numerical Prandtl number and its resolution dependency in MRI simulations is clear, as a low-order resistive scheme with a high-order viscosity scheme will eventually result in a low $P_m$ value and can lead to misinterpretation in convergence studies. The numerical Prandtl number has been investigated in several grid codes (\cite{2007A&A...476.1123F}[{with \sc ZEUS}] \cite{2007MNRAS.381..319L}[{with \sc ZEUS3D}] \cite{2009ApJ...690..974S}[{with \sc ATHENA}] \cite{2011PhRvL.107k4504F}[{with \sc FLASH}]), which have found a Prandtl number of around $P_m \sim 2$ with a very weak dependency on resolution. However, the true $P_m$ value during non-linear MRI simulation remains uncertain as the numerical dissipation is not readily available for grid codes and requires comparison to analytical work or analysis of Fourier transfer functions with certain assumptions/constraints. The estimates of the numerical Prandtl number in these papers are taken for subsonic flows and might significantly change for higher Mach flows. In this paper, we will take a closer look at the numerical Prandtl number in SPH and see how it affects the turbulence within MRI simulations. 
\\ \\
The vast majority of MRI simulations have been carried out with Eulerian grid-based codes. There have only been a handful of studies investigating the MRI with meshless methods in 2D \citep{2011MNRAS.414..129G,2013MNRAS.432..176P,2016MNRAS.455...51H} and in 3D \citep{2019ApJS..241...26D}. The MRI is an especially difficult test for meshless codes due to the strong divergence-free constraint, which in Eulerian codes can be enforced to machine precision with the constrained transport method \citep{1988ApJ...332..659E}. However, improved divergence cleaning methods in recent years have been developed for meshless codes, which significantly reduce the divergence errors \citep{2012JCoPh.231.7214T,2016JCoPh.322..326T}. A benefit of Lagrangian methods such as SPH is that they are always Galilean invariant and do not suffer from advection errors, which can otherwise be an issue for Eulerian codes in simulations with large bulk flows. In addition, SPH is naturally adaptive in resolution, making it ideal for simulations involving a wide range of spatial scales. Understanding the numerical aspects of the MRI in SPH is important, as SPH is widely used in astrophysical simulations where the MRI can be present.  
\\ \\
In \cite{2019arXiv190105190D} the authors investigated the MRI in 3D with the meshless finite mass (MFM) and the SPH methods for a wide array of different initial magnetic field configurations. For the unstratified NF case, it was shown that MFM and SPH showed similar behavior to Eulerian grid-based codes. However, for the unstratified ZNF case, both MFM and SPH showed rapid decay of the turbulence. This is likely related to the numerical dissipation schemes of the two methods, which we will investigate for SPH in this paper. For the stratified azimuthal NF case, the MFM method could correctly produce the characteristic dynamo cycles for around 50 to 70 orbits before the turbulence eventually died out. SPH on the other hand could not develop sustained turbulence and instead developed unphysically strong azimuthal fields. This was attributed to a combination of discretization errors of the magnetic field in the radial component and divergence cleaning amplifying the vertical field component. In this paper, we will further investigate this case with the newly developed Geometric Density SPH (GDSPH), which has been shown to improve the accuracy of SPH in problems involving large density gradients. Specifically, it allows grid-scale instabilities to grow that are suppressed in traditional SPH  \citep{2017MNRAS.471.2357W,2020A&A...638A.140W}.
\\ \\
In this paper, we have performed MRI simulations of the unstratified NF case in the regular-sized box ($L_{z}/L_{x}=1$) and the unstratified ZNF case in both regular and taller sized boxes ($L_{z}/L_{x}=4$ as in \cite{2016MNRAS.456.2273S}) with varying resolution and numerical dissipation parameters. We have investigated the numerical Prandtl number in SPH and its effect on the amplification and saturation of the MRI. We have also performed simulations on the stratified NF case with both the traditional SPH method (TSPH) and the GDSPH method to further investigate the unphysical growth in the azimuthal fields observed in \cite{2019arXiv190105190D}. For these simulations, we also vary the resolution and strength of the numerical dissipation. In addition, to all the simulations we also investigate the turbulent transport coefficients. 
\\ \\
This paper is organized as follows. In Section~\ref{sec:theory}, we go through the basics of dynamo theory, the simulation setup, and the post-process analysis. In Section~\ref{sec:unstratresult}, we present our result for the unstratified NF and ZNF cases and in Section~\ref{sec:SNF} we present our result for the stratified case. In Section~\ref{sec:discussion}, we discuss our results and present some concluding remarks.
\section{Theory}
\label{sec:theory}
\subsection{Dynamo theory}
\label{subsec:dynamo}
A magnetic dynamo describes the exponential growth and sustenance of magnetic fields due to being stretched, twisted, and folded by the underlying fluid motions. While specific velocity field configurations can lead to dynamo action (laminar dynamo), astrophysical fluids are usually highly turbulent, where motion is chaotic across a large range of spatial scales. Dynamo action can occur across all turbulent scales, but the magnetic field is stretched faster by the smaller scale motions than the larger-scale ones, leading to a faster growth on smaller scales (known as the small-scale dynamo) \citep{1992ApJ...396..606K,1999ARA&A..37...37K}. In ideal MHD, the growth rate is set primarily by the viscous scale of the fluid. However, for MHD with diffusion of the magnetic field (resistivity), this is no longer necessarily true, as the magnetic fields on small-scales can now be damped quickly. This makes the growth of the magnetic field more intricate, as it is determined by the relationship between the viscous scale $l_v$ and the resistive scale $l_{\eta}$ \citep{1962pfig.book.....S}. The ratio between these two, the magnetic Prandtl number $P_m=\frac{l_v}{l_{\eta}}$ is thus very important in the resulting characteristic and saturation of the turbulent dynamo \citep{2004ApJ...612..276S}. For $P_m>1$ the quickest twisting and folding of the magnetic field is driven at the viscous scale, where the underlying velocity field is smooth as there are no smaller velocity structures in the flow at this scale. The chaotic but smooth motion at this scale lends itself to dynamo action, which means that magnetic fields can efficiently be generated \citep{1972SvPhU..15..159V,1983mfa..book.....Z,1990alch.book.....Z}. The cut-off scale of magnetic fluctuations will still be set by the resistive scale, which allows for a buildup of power within the subviscous range. With higher $P_m$ values than one, more of the subviscous scale becomes available for magnetic field amplification \citep{1992ApJ...396..606K,2004ApJ...612..276S}. 
\\ \\
The small-scale dynamo has been shown to be a possible mechanism for amplifying the weak seed fields in the early universe to magnitudes observed in galaxies today \citep{1990ApJ...365..544B,1996ARA&A..34..155B,1997ApJ...480..481K}. However, magnetic fields in the universe exhibit a high degree of coherence at scales larger than the underlying turbulent motion \citep{2005A&A...444..739B,2015A&ARv..24....4B}. Large-scale dynamo theory is an attempt to explain how these coherent large-scale magnetic field structures can be generated in highly turbulent environments. In essence, it investigates how the small-scale kinetic and magnetic fluctuations couple to the underlying large-scale field. 
\\ \\
To figure out the effect of the small-scale field on the large-scale field, it is useful to introduce the formalism of mean-field theory \citep{1978mfge.book.....M,1979cmft.book.....P,1980opp..bookR....K,1988ASSL..133.....R,2005PhR...417....1B}. Assuming a scale separation between the large-scale and small-scale, both the magnetic and velocity fields can be decomposed to a mean field component ($\overline{B}$ and $\overline{U}$ ) and a fluctuating component ($b$ and $u$):
\begin{equation}
\label{eq:meanfieldparts}
B=\overline{B}+b \quad U=\overline{U}+u
\end{equation}
Averaging the induction equation leads to the evolution equation for the magnetic mean-field: 
\begin{equation}
\label{eq:meanfieldinduc}
\frac{\partial\overline{B}}{\partial t}=\nabla \times (\overline{U} \times \overline{B}) + \nabla \times \mathcal{E} + \eta \nabla^2 \overline{B} 
\end{equation}
Here $U$ represents the large-scale velocity structure, $\eta$ the magnetic diffusivity and $\mathcal{E}$ is the electromotive force (EMF) produced by the fluctuating fields.
\begin{equation}
\label{eq:smallscaleemf}
\mathcal{E}=\overline{u \times b}
\end{equation}
By studying how the fluctuating $u$ and $b$ fields reacts to an applied mean field, it can be shown that both $u$ and $b$ contain a component independent of the mean-field and an additional term which is linearly dependent on the applied mean-field.
\begin{equation}
\label{eq:turbfield}
b=b_0+b_{\overline{B}} \quad u=u_0+u_{\overline{B}}
\end{equation}
Assuming the indepentent terms $b_0$ and $u_0$ are uncorrelated($\mathcal{E}_0=\overline{u_0 \times b_0}=0$) and the assumption of scale separation, we can expand $\mathcal{E}$ in a Taylor series in $\overline{B}$ and $\overline{U}$:
\begin{equation}
\label{eq:eps}
\mathcal{E}_i=\alpha_{ij}\overline{B_j}-\eta_{ij}\overline{J_j}+\gamma_{ij}\overline{\Omega_j}+...
\end{equation}
Here $\alpha$, $\eta$ and $\gamma$ are the tensorial transport coefficients and $\overline{J}=\overline{\nabla\times B}$ is the mean-field current density and $\overline{\Omega}=\overline{\nabla\times U}$ is the mean-field vorticity. The first term of \eq~\ref{eq:eps} is the $\alpha$-effect in which the small-scale turbulence generates an EMF which is proportional to the mean-field itself. This effect, coupled together with differential rotation, can develop the well-known $\alpha\omega$ dynamo. The alpha-effect depends crucially on the small-scale helcities within the turbulent flow, which require the system to break statistical symmetry either by stratification or through having a net helicity \citep{1976JFM....77..321P,1978mfge.book.....M,2005PhR...417....1B}. The second term in \eq~\ref{eq:eps} generates an EMF in proportion to the mean-current and can act to either amplify or diffuse the mean-field. The last term of \eq~\ref{eq:eps} is the Yoshizawa effect, which acts without the need for a large-scale magnetic field and can be seen as a turbulent battery mechanism \citep{1993ApJ...407..540Y,2013GApFD.107..114Y}. In addition to a mean vorticial velocity component, the effect requires small-scale cross-helicity between the turbulent fields ($\overline{u \cdot b}$). \footnote{A global magnetic field is probably required to create cross-helicity in a turbulent field. Cross-helicity is shown to occur when the mean-field is parallel to the direction of gravity in \cite{2011SoPh..269....3R}}
\\ \\
In this paper, we are interested in the dynamo action that arises within shearing boxes (see section~\ref{subsec:simsetup} for coordinate definitions and simulation setup). We define our mean-field by taking a horizontal average:
\begin{equation}
\label{eq:horzavg}
\overline{X}=\frac{\int X dx dy }{\int dx dy}
\end{equation}
The turbulent field can then be calculated by removing the mean-field component from the total field (see \eq~\ref{eq:meanfieldparts}). For the velocity, the mean-field is determined from the shearing box approximation ($U_0=-q\Omega x \hat{\mathbf{y}}$). Here $\Omega$ is the angular velocity and $q$ is the shearing parameter, which for Keplerian disks is $q=3/2$.  Since the horizontal average is only a function of $z$, \eq~\ref{eq:eps} and subsequently \eq~\ref{eq:meanfieldinduc} simplifies greatly ($\overline{B_z}=0$, $\overline{J_z}=0$, and $\overline{\Omega_z}=0$):
\begin{equation}
\label{eq:emfx}
\mathcal{E}_x=\alpha_{xx}\Bmx+\alpha_{xy}\Bmy-\eta_{xx}\Jmx-\eta_{xy}\Jmy
\end{equation}
\begin{equation}
\label{eq:emfy}
\mathcal{E}_y=\alpha_{yx}\Bmx+\alpha_{yy}\Bmy-\eta_{yx}\Jmx-\eta_{yy}\Jmy
\end{equation}
\begin{equation}
\label{eq:dBdtx}
\frac{\partial\overline{B}}{\partial t}_x=-\partial_z(\alpha_{yx}\Bmx)-\partial_z(\alpha_{yy}\Bmy)+\partial_z(\eta_{yx}\Jmx)+\partial_z((\eta_{yy}+\eta)\Jmy)
\end{equation}
\begin{equation}
\label{eq:dBdty}
\frac{\partial\overline{B}}{\partial t}_y=-q\Bmx+\partial_z(\alpha_{xx}\Bmx)+\partial_z(\alpha_{xy}\Bmy)-\partial_z((\eta_{xx}+\eta)\Jmx)-\partial_z(\eta_{xy}\Jmy)
\end{equation}
The $\gamma$ terms from \eq~\ref{eq:eps} become zero as our only component of $\overline{\Omega}$ is in the z direction. The diagonal components $\alpha_{xx}$ and $\alpha_{yy}$ are the main driver of the alpha effect, generating a feedback loop between the radial and azimuthal fields. The sign of diagonal components will depend on $\Omega \cdot g$ (here $g$ is the gravitational acceleration) which will give us an odd symmetry around the midplane of our stratified box simulations. Depending on the gradient of the $\alpha$  parameter and structure of the magnetic field, it will either work in accordance or in discordance with the field. The anti-symmetric components $\alpha_{xy}$ and $\alpha_{yx}$ can be related to the diamagnetic pumping term $\gamma_z=\frac{1}{2}(\alpha_{yx}-\alpha_{xy})$ and describes the transport of mean-fields due to the turbulence. In a similar fashion, the diagonal components $\eta_{xx}$ and $\eta_{yy}$ describe the diffusion of the mean-field due to the turbulence. Finally, we have the off-diagonal components $\eta_{xy}$ and $\eta_{yx}$ that are responsible for the dynamo produced by the $\Omega \times J $ effect \citep{1969WisBB..11..194R} and the shear current effect \citep{2003PhRvE..68c6301R}. 
These equations provide a powerful tool to connect our simulation data to the mean-field theory. To calculate the transport coefficients we can fit \eq~\ref{eq:emfx} and \eq~\ref{eq:emfy} to output data from our simulations. One starts by calculating
\begin{equation}
\label{eq:matrix1}
A_{1} = \left[ \Bmx\mathcal{E}_x, \Bmy\mathcal{E}_x,\Jmx\mathcal{E}_x, \Jmy\mathcal{E}_x \right ]
\end{equation}
and 
\begin{equation}
A_{2} = \left[ \Bmx\mathcal{E}_y, \Bmy\mathcal{E}_y,\Jmx\mathcal{E}_y, \Jmy\mathcal{E}_y  \right] 
\end{equation}
and the matrix
\begin{equation}
\label{eq:matrix2}
M=
\begin{pmatrix}
\Bmx\Bmx & \Bmx\Bmy & \Bmx\Jmx & \Bmx\Jmy\\
\Bmy\Bmx & \Bmy\Bmy & \Bmy\Jmx & \Bmy\Jmy\\
\Jmx\Bmx & \Jmx\Bmy & \Jmx\Jmx & \Jmx\Jmy\\
\Jmy\Bmx & \Jmy\Bmy & \Jmy\Jmx & \Jmy\Jmy\\
\end{pmatrix}
\end{equation}
Then we solve the following matrix equations (using a least-square method):
\begin{equation}
\label{eq:matrix3}
    A_{1} = MC_{1} \ {\rm and} \ A_{2} = MC_{2}
\end{equation}
for the transport coefficients: 
\begin{equation}
\label{eq:matrixsol}
C_{1} = (\alpha_{xx}, \alpha_{xy}, -\eta_{xx}, -\eta_{xy}) \ {\rm and} \
C_{2} = (\alpha_{yx}, \alpha_{yy}, -\eta_{yx}, -\eta_{yy})
\end{equation}
Because the mean-field is not solely evolved by $\mathcal{E}$ but also by the shearing and dissipation of the field, significant errors can arise in the transport coefficients from the correlation between different components. The main harmful error comes from correlations with $\Bmx$(due to the shear term). We can improve the signal and reduce the noise by minimizing the influence of $\Bmx$ with the following two approximations. First, we set the diagonal transport coefficients to be equal ($\alpha_{xx}=\alpha_{yy}$ $\eta_{xx}=\eta_{yy}$), which have been shown to be an accurate approximation \citep{2009MNRAS.398.1891H,2010MNRAS.405...41G}. The second approximation is to set $\alpha_{yx}=0$, $\eta_{xy}=0$ which is justified by the fact that $\Bmx<<\Bmy$ \citep{2015PhRvL.115q5003S}. However, we have seen that including $\alpha_{yx}$, $\eta_{xy}$ does not significantly change the result for the other transport coefficients. For comparison with previous studies, in the stratified case we allow a non-zero $\alpha_{yx}$ and $\eta_{xy}$, and allow $\alpha_{xx}$ and $\alpha_{yy}$ to be different.
\\ \\
What causes the dynamo growth within shearing box simulations of the MRI remains uncertain and remains an active area of research. The unstratified shearing box simulations develop a so-called non-helical shear dynamo which cannot be generated by the $\alpha$-effect (in the traditional sense) as there is no net kinetic/magnetic helicity or density stratification within the flow. This implies that the mean of the $\alpha$ coefficients will tend towards zero. However, the $\alpha$ coefficients for a finite-sized system will fluctuate in time. If the fluctuations are sufficiently large, this has shown to enable dynamo growth. This has been called the incoherent-$\alpha$ dynamo, which could explain the dynamo mechanism in unstratified shearing boxes \citep{1997ApJ...475..263V,2000A&A...364..339S,2011PhRvL.107y5004H}. However, a potential issue with the incoherent-$\alpha$ dynamo is that the mean fluctuations in $\alpha$ becomes smaller as the size of the box is increased, which decreases the growth rate of the dynamo. Another potential mechanism for the dynamo is the magnetic shear-current effect, which depends crucially on the off-diagonal turbulent resistivity coefficient $\eta_{yx}$ \citep{2003PhRvE..68c6301R,2015PhRvE..92e3101S,2015PhRvL.115q5003S,2015PhRvL.114h5002S}. The idea of the magnetic shear-current effect is that a bath of magnetic fluctuations under the influence of an azimuthal field produces an EMF that generates radial fields which subsequently act to amplify the azimuthal field resulting in a dynamo instability. The instability can be shown to happen if $-\eta_{yx}\Omega_z<0$ which means that dynamo action is possible from the shear-current effect if $\eta_{yx}$ is negative. In this paper, we will examine the shear-current effect and determine if it agrees well with previous results.
\\ \\
Furthermore, stratification adds several additional mechanisms that can affect the dynamo process. The $\alpha$-effect can provide dynamo action and has been proposed to be a main driver of the dynamo together with the shear-current effect. Both of these effects can cause a phase-shift between the growing fields, explaining the cyclic nature of the radial and azimuthal fields seen in stratified shearing boxes. In addition, buoyancy instabilities expel the magnetic field outwards.

\subsection{Simulation setup}
\label{subsec:simsetup}
For all our simulations we use the MHD version of {\sc Gasoline2} with the same default set of code parameters as in \cite{2020A&A...638A.140W}. The simulations are set up using a shearing box approximation \citep{1965MNRAS.130..125G,1995ApJ...440..742H}, in which a co-rotating patch of disk with angular velocity $\Omega$ at a distance $R$ is used as the computational domain. The patch is assumed small such that curvature can be neglected and we can employ a local cartesian coordinate system with $x$ as the radial direction, $y$ as the azimuthal direction and $z$ as the vertical direction. The additional terms added to the equations of motion are:
\begin{equation}
    \left(\frac{dv}{dt}\right)_{shearbox}=2q\Omega^2x \ \mathbf{\hat{x}}-2 \mathbf{\Omega} \times \mathbf{v} - \Omega^2 z \ \mathbf{\hat{z}}
    \label{eq:shearapprox}
\end{equation}
$$ q = -\frac{d\ln \Omega}{d\ln r} $$
Where the term $2q\Omega^2x \ \mathbf{\hat{x}}$ represents the tidal acceleration, $2 \mathbf{\Omega} \times \mathbf{v}$ represents the Coriolis force and $\Omega^2 z \ \mathbf{\hat{z}}$ represents the vertical gravitational force from the central object. The equilibrium solution of \eq~\ref{eq:shearapprox} will be independent of time and follows a uniform shearing motion in the azimuthal direction
\begin{equation}
\mathbf{v}_y=q\Omega x \ \hat{\mathbf{y}}.
\end{equation}{}
If any mean radial velocity exists, the simulation box will start to oscillate with an epicyclic frequency of $\kappa=2\Omega\sqrt{1-q/2}$. This can be the case if set initially or if momentum is not tightly conserved. SPH conserves both momentum and energy very tightly and only suffers from non-conservation in the strong-field regime ($\beta<2$) in the presence of large divergence errors. The shear parameter $q$ is set to follow a Keplerian profile ($q=3/2$) and the angular velocity is set to $\Omega=1.0$. The boundary in the x-direction is shear periodic, which means that particles passing/interacting across the boundary receives a velocity offset of $\Delta v=\pm q\Omega L_x \hat{\mathbf{y}}$ in which $L_x$ is the length of the domain in the $x$ direction. This is simpler than for grid codes, where shear periodic boundaries require careful reconstruction to retain conservative fluxes across the boundary. While retaining fluxes due to boundary conditions remain simple in SPH, there are other potential flux errors. The main one comes from the divergence error and, more precisely, the removal of the monopole current $v(\nabla\cdot B)$ from the induction equation\citep{2000JCoPh.160..649J,2001JCoPh.172..392D,2005MNRAS.364..384P}. Removing this term from the induction equation ensures that the surface flux is conserved (which is crucial) and makes the magnetic field divergence become a passive scalar which is simply carried away with the flow. However, in doing this, it is no longer ensured that the volume integral of the magnetic field is conserved(the global mean-field). This issue is shared among the majority of numerical schemes and in this case the error will directly depend on the magnetic field divergence. The error is generally very small but can contaminate the solution, as MRI can be quite sensitive to any global radial mean-field. To avoid this we employ a correction to the flux, which ensures that no global radial mean-fields are generated. We do not employ this correction for simulations with outflow boundaries.
\\ \\
For the unstratified simulations, the last term of \eq~\ref{eq:shearapprox} is not included and the domain is periodic in both the y and z directions. For the stratified simulations, all terms are included and the domain is periodic in y and has outflow boundaries in z. The outflow boundary in z is set to remove any element with a smoothing length greater than $h=0.5 L_x$ to avoid the double-counting of elements across the computational domain. The stratified simulations acquire a density profile of $\rho=e^{-zH}$ with a scale height of $H=c_s/\Omega$ where $c_s$ is the speed of sound. We use an isothermal equation of state ($P=\rho c_s^2$), with $c_s=1.0$ set for all simulations. Before simulations are run, the initial particle distribution is relaxed to a glass distribution, then random velocity perturbations of around $5\%$ of the sound speed are added to the shear flow to quickly initiate the MRI.
\\ \\
To determine how well the MRI is resolved, we use the resolution metric developed by \cite{2010ApJ...711..959N} which defines an effective quality parameter (number of resolution elements per MRI wavelength):
\begin{equation}
\label{eq:quality}
    Q=\frac{\lambda_{MRI}}{h}=\frac{2\pi v_{a,z}}{\Omega h},
\end{equation}
where $\lambda_{MRI}$ is the characteristic wavelength and is roughly equal to the fastest growing MRI mode, and $v_{a,z}$ is the vertical component of the Alfvén velocity, and $h$ is the resolution length. We follow the example from \cite{2019ApJS..241...26D} where, instead of setting the resolution length to the smoothing length, it is based on the standard deviation of the smoothing kernel. For the Wendland C4 kernel, it gives an effective resolution element length $h_{eff} = 0.9h$. To properly resolve the linear MRI roughly only $Q>6$ is required, however, the stress is highly resolution-dependent until a value of roughly $Q_z>10$ and $Q_y>20$ is reached (for the stratified NF case \citet{2011ApJ...738...84H}).
\\ \\
As mentioned in the introduction, the magnetic Prandtl number plays a large role in the growth of turbulent dynamos. To gauge the effective Prandtl number from a numerical scheme, the numerical dissipation needs to be determined and translated into an effective kinematic viscosity ($\nu$) and physical resistivity ($\eta$). In Eulerian schemes, dissipation comes partly from advection of the fluid which introduces diffusion due to truncation errors in the flux reconstruction. This error will be proportional to both the resolution and the fluid velocity. For shear periodic boundary conditions this means that there will be uneven dissipation due to larger velocities near the edge than the centre of domain. Moreover, additional dissipation is added to maintain numerical stability, this is often done through Riemann solvers in grid codes. Estimating the numerical diffusion in Eulerian schemes is not straightforward and often requires comparison to analytical solutions for an accurate estimate.
\\ \\
Compared to Eulerian grid codes, SPH does not suffer from these advection errors and artificial dissipation terms\footnote{The artificial dissipation terms can be seen as approximate Riemann solvers as they functionally produce similar dissipation \citep{1997JCoPh.136..298M} } are added to handle flow discontinuities (e.g. shocks). These are primarily discretized from physical dissipation laws, but with diffusion parameters that depend on the resolution and potentially on flow properties. In \cite{1985CoPhR...3...71M} it was shown that the linear coefficient($\alpha_{AV}$) in the artificial viscosity corresponds to a resolution-dependent physical viscosity in the continuum limit, which has been confirmed by several authors \citep{1994ApJ...421..651A,2010MNRAS.405.1212L,2012MNRAS.427.2022M}.
However, extrapolating to the continuum limit in this case underestimates the physical viscosity/resistivity. It becomes more difficult to estimate the physical dissipation when using particle-pair dependent signal velocities (as is done for our artificial resistivity). In this paper, we opt for another way to estimate the physical dissipation. By recording the energy lost due to artificial dissipation terms, we can directly estimate the parameters from the equivalent physical dissipation equations. From the Navier-Stokes equation we can estimate the shear viscosity with:
\begin{equation}
\label{eq:shearvisc}
\nu_{AD}=\frac{\left(\frac{du}{dt}\right)_{AV}}{\frac{1}{2}\left(\frac{\partial v^i}{\partial x^j}+\frac{\partial v^j}{\partial x^i}\right)^2+(\nabla\cdot\mathbf{v})^2}
\end{equation}{}
Here, we have assumed the fixed ratio between the bulk viscosity and the shear viscosity, which follows from the continuum limit derivation ($\zeta_{AV}=\frac{5}{3}\nu_{AV}$) \citep{2010MNRAS.405.1212L}. We estimate the physical resistivity from the Ohmic dissipation law:
\begin{equation}
\eta_{AD}=\frac{\rho}{J^2}\left(\frac{du}{dt}\right)_{AR}
\end{equation}{}
Taking the ratio of the two equations then gives us the numerical Prandtl number:
\begin{equation}
P_{m,AD}=\frac{\nu_{AD}}{\eta_{AD}}
\end{equation}{}
For some of our simulations we force a certain average numerical Prandtl number $\Vm{P_{m,AD}}=\frac{\Vm{\nu_{AD}}}{\Vm{\eta_{AD}}}$. This is done by adjusting either $\left(\frac{du}{dt}\right)_{AV}$ and $\left(\frac{dv}{dt}\right)_{AV}$ or $\left(\frac{du}{dt}\right)_{AR}$ and $\left(\frac{dB}{dt}\right)_{AR}$ by a constant factor such that $\Vm{P_{m,AD}}$ corresponds to the desired value. This is the same as changing the artificial dissipation coefficients $(\alpha_B,\alpha_{AV})$ by a constant factor at each time step.

\begin{figure*}[!h]
    \centering
    \includegraphics[width=\hsize]{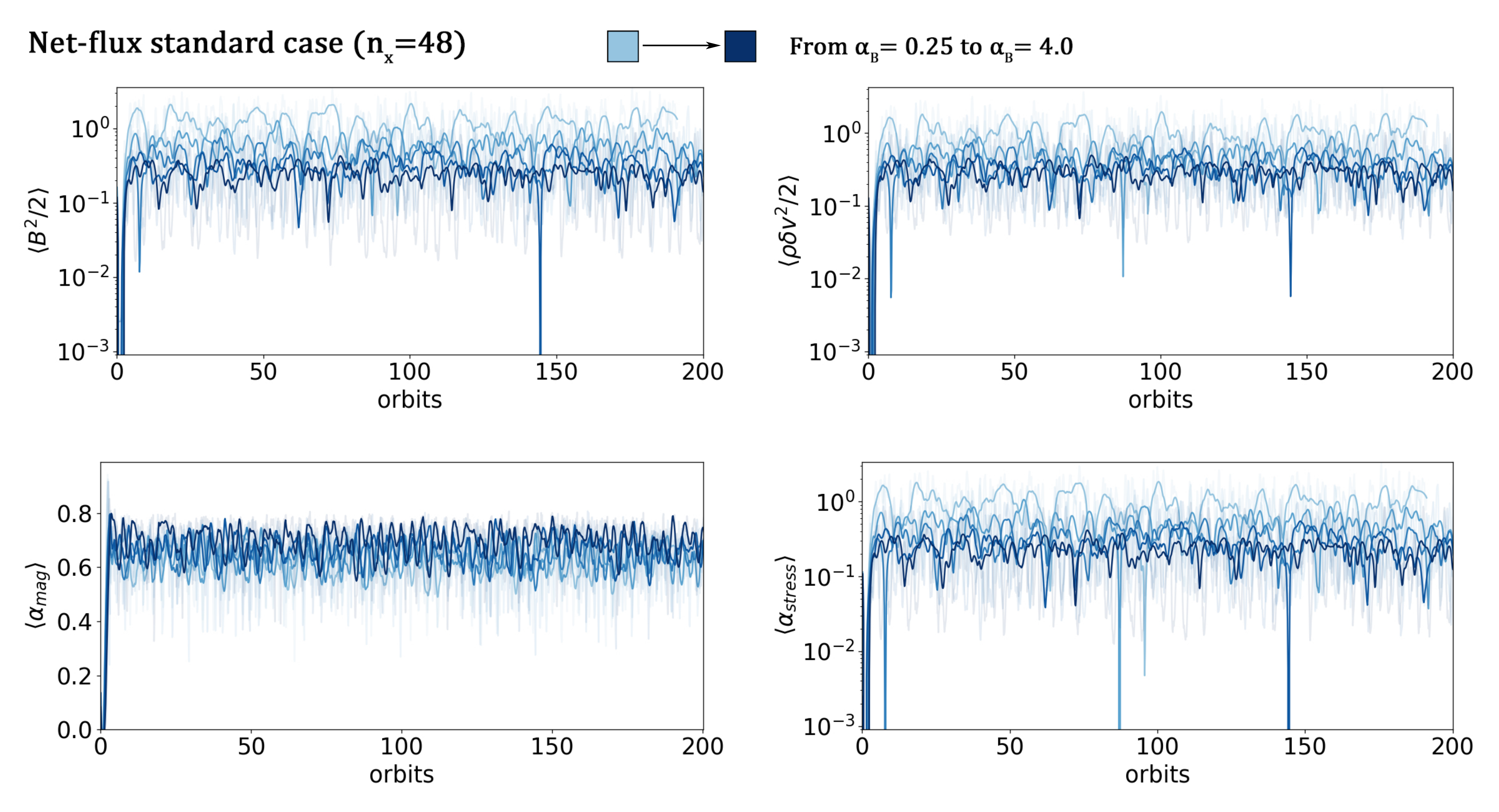}
    \caption{The time evolution of several volume-averaged quantities over 200 orbits. Magnetic energy(top left), kinetic energy(top right), normalized Maxwell stress(bottom left) and the total stress(bottom right). The darkness of the curves is determined by the strength of the artificial resistivity parameter, $\alpha_B=0.25,0.5,1.0,2.0,4.0$.
    Due to the high oscillatory nature of the simulation we have smoothed the curves using a Savitzky–Golay filter, the unsmoothed curves can still be seen as very transparent curves. The oscillations are related to the formation and destruction of channel modes.}
    \label{fig:UNFtimeevo}
\end{figure*}
\begin{figure*}[!h]
    \centering
    \includegraphics[width=\hsize]{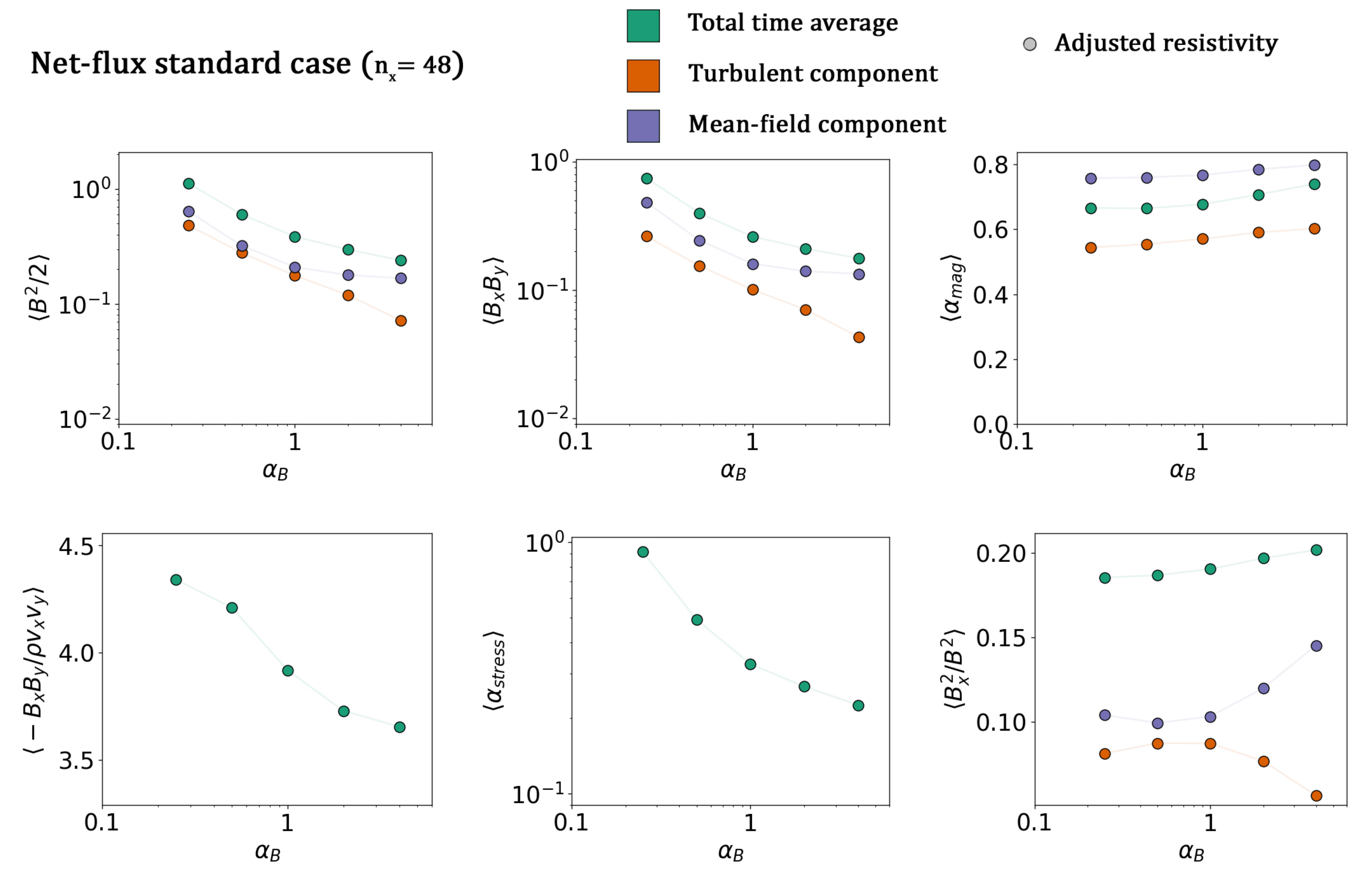}
    \caption{Time-averaged values of several quantities as a function of the artificial resistivity coefficient, for all our unstratified net-flux simulations. From the top left to bottom right, we plot the magnetic energy density, the Maxwell stress, the normalized Maxwell stress, the ratio between Reynolds and Maxwell stresses, the total stress, the ratio between radial and total magnetic field energy. For some quantities we have plotted the total time average (shown in green), the time average of the turbulent component(shown in orange) and the time average of the mean-field component(shown in blue).}
    \label{fig:UNFaverage}
\end{figure*}
\subsection{Post-process analysis}
\label{postanal}
After the SPH simulations are done, the particle data is interpolated to uniform grid data for post-analysis. To obtain statistical properties of our simulations, we average our data in a few different ways. The first is the horizontal average given in \eq~\ref{eq:horzavg}. The second is the volume average:
\begin{equation}
\label{eq:volavg}
\langle X \rangle=\frac{\int X dV }{\int dV}
\end{equation}
and the final one is the time average:
\begin{equation}
\label{eq:timeavg}
\langle X \rangle_t =\frac{\int X dt }{\int dt}
\end{equation}
All the averages are in general applied over the whole spatial/time domain of the simulation if not stated otherwise. To quantify the angular momentum transport and the saturation of the MRI, it is useful to calculate the stresses in the fluid. The total stress together with its magnetic and hydrodynamic component is given by:
\begin{equation}
    \alpha_{stress} = -\frac{B_xB_y}{P_0}+\frac{\rho (v_x-\Bar{v_x})(v_y-\Bar{v_y})}{P_0},
\end{equation}
where $P_0$ is the initial pressure(in our case $P_0=1$) and $\rho$ is the density and here the first term in the equation represent the Maxwell stress ($\alpha_{MW}$) and the second term the Reynolds stress ($\alpha_{Rey}$). A related quantity that we look at is the normalized magnetic stress:
\begin{equation}
    \alpha_{mag}=-2\frac{\Vm{B_xB_y}}{\Vm{B^2}}
\end{equation}
In addition to looking at the effect of the total field, we will also investigate the contributions from the mean-field ($\Bm$) and turbulent component (\textbf{b}) in the magnetic energy and the stress. We define their respective normalized stress as in \citet{2016MNRAS.456.2273S}:
\begin{equation}
    \alpha_{mag,mean}=-2\frac{\Vm{\Bmx\Bmy}}{\Vm{\Bm^2}} \quad
    \alpha_{mag,turb}=-2\frac{\Vm{b_xb_y}}{\Vm{B^2-\Hm{B^2}}}
\end{equation}
Another useful quantity is the Elsasser number, which describes the relative strength of the magnetic dissipation term:
\begin{equation}
\Lambda = \frac{v_a^2}{\eta_{AD} \Omega}
\end{equation}
Here, $v_a$ is the Alfven speed. For a $\Lambda<1$ the linear properties of the MRI will change significantly and hinder saturation \citep{1994ApJ...421..163B,1999MNRAS.307..849W,2001ApJ...552..235B}.
\\ \\
Finally, it is important to track the divergence error in numerical simulations to make sure it remains small and does not severely effect the results.
\begin{equation}
\label{eq:divBerr}
\epsilon_{divB}=\frac{h|\nabla\cdot \Bv|}{|B|} .
\end{equation}
The mean of this quantity should preferably remain below $10^{-2}$ but higher values can still be acceptable (depending on the system).

\section{Unstratified simulation results}
\label{sec:unstratresult}

\subsection{Net-flux simulations}
\label{subsec:UNF}
\begin{figure}[!h]
    \centering
    \includegraphics[width=4cm]{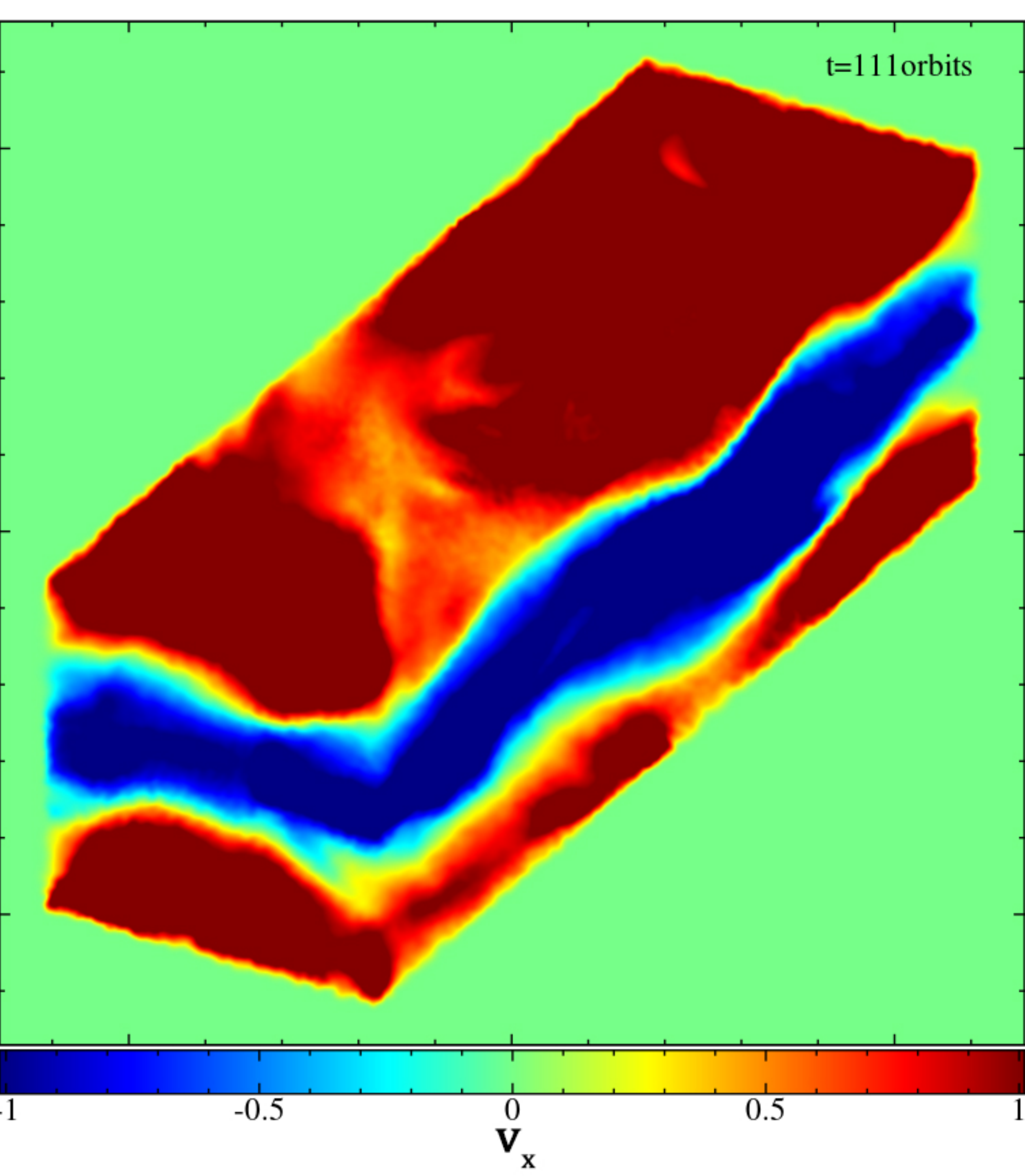}
    \includegraphics[width=4cm]{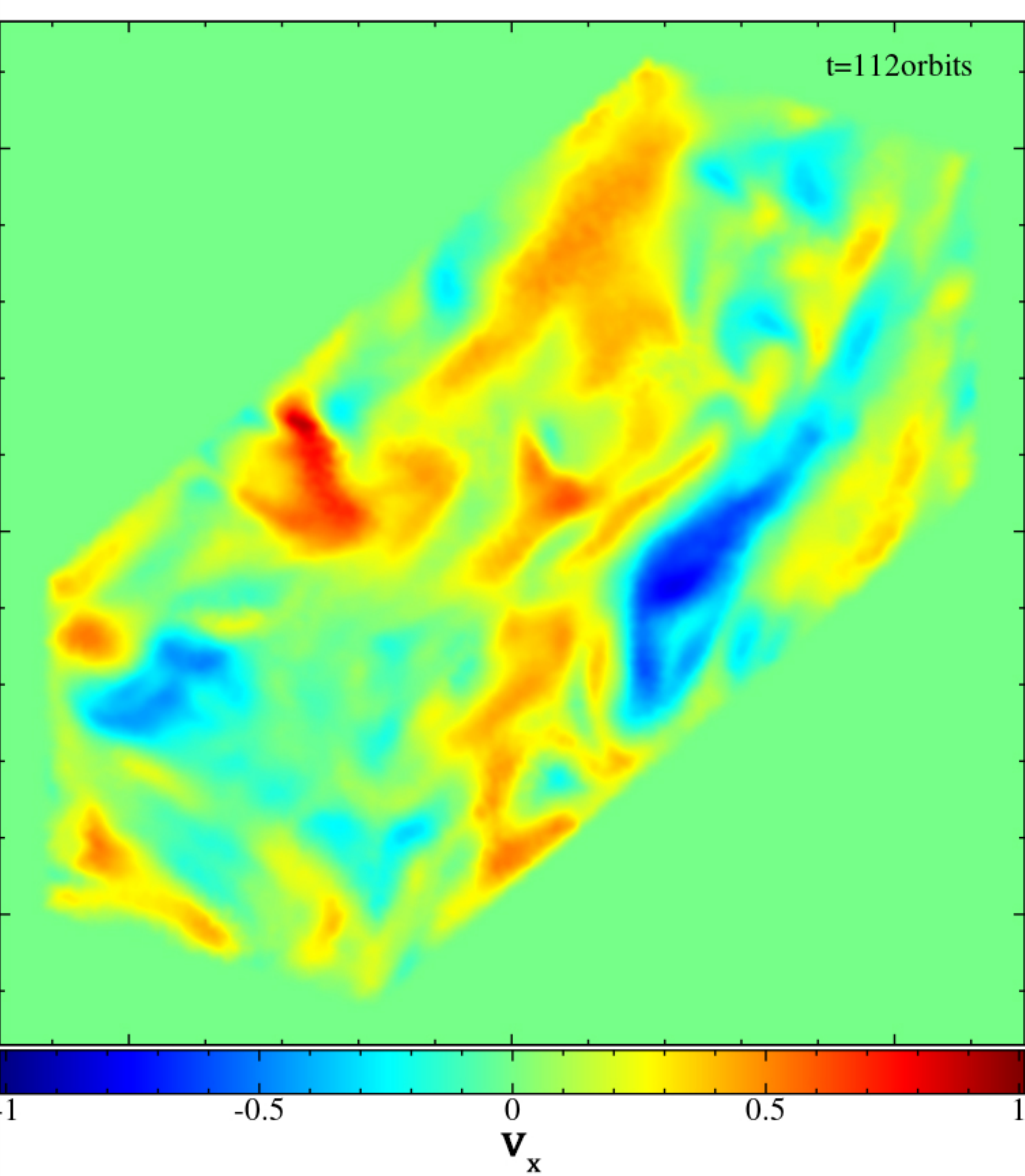}
    \caption{ The generation and break down of a channel mode during the unstratified net-flux run with $\alpha_B=0.5$. The figure depicts a surface rendering of the radial velocity within the shearing box. The two-channel flow is clearly seen in the left picture which over an orbit is quickly broken down into turbulence, as seen in the right figure. The generation of the channel mode coincides with a peak in the magnetic energy and as the channel flow is destroyed the magnetic energy will decrease. The formation and destruction of these channel flows occur continuously throughout the simulation. }
    \label{fig:UNFchannelmode}
\end{figure}
\begin{figure*}[!h]
    \centering
    Net-flux case $\alpha_{B}=0.5$ \qquad \qquad \qquad \qquad \qquad \qquad \qquad \qquad \qquad Net-flux case $\alpha_{B}=4$\\
    \includegraphics[width=9cm]{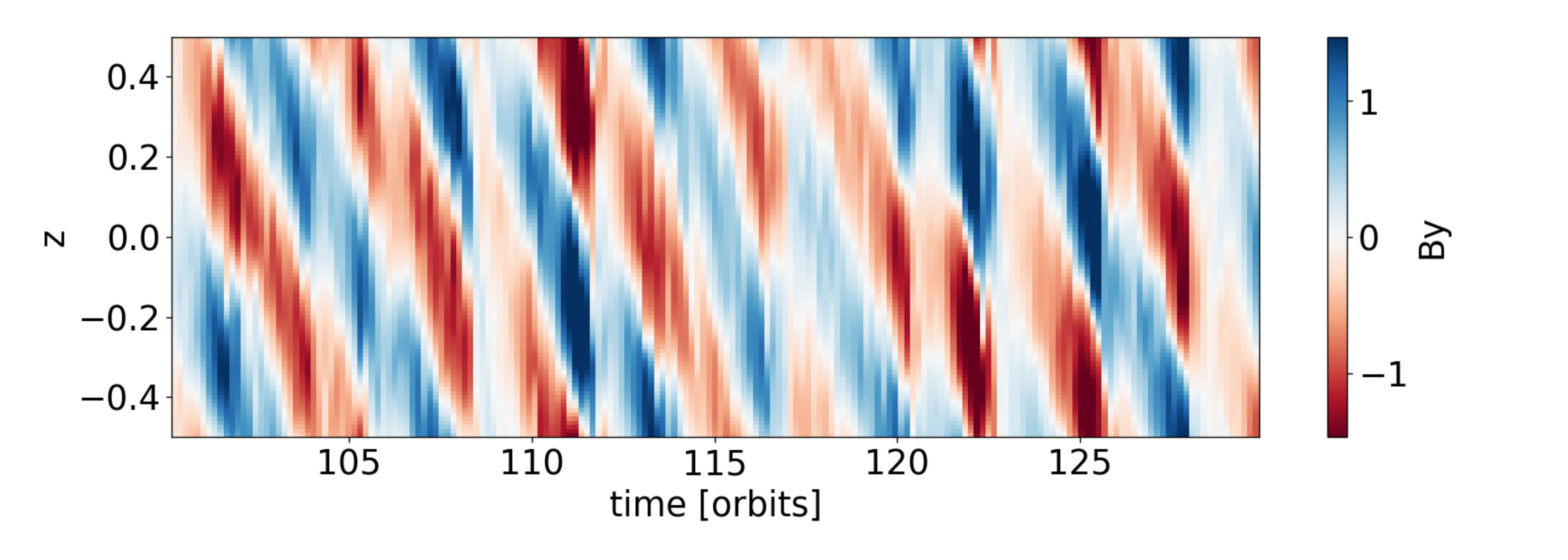}
    \includegraphics[width=9cm]{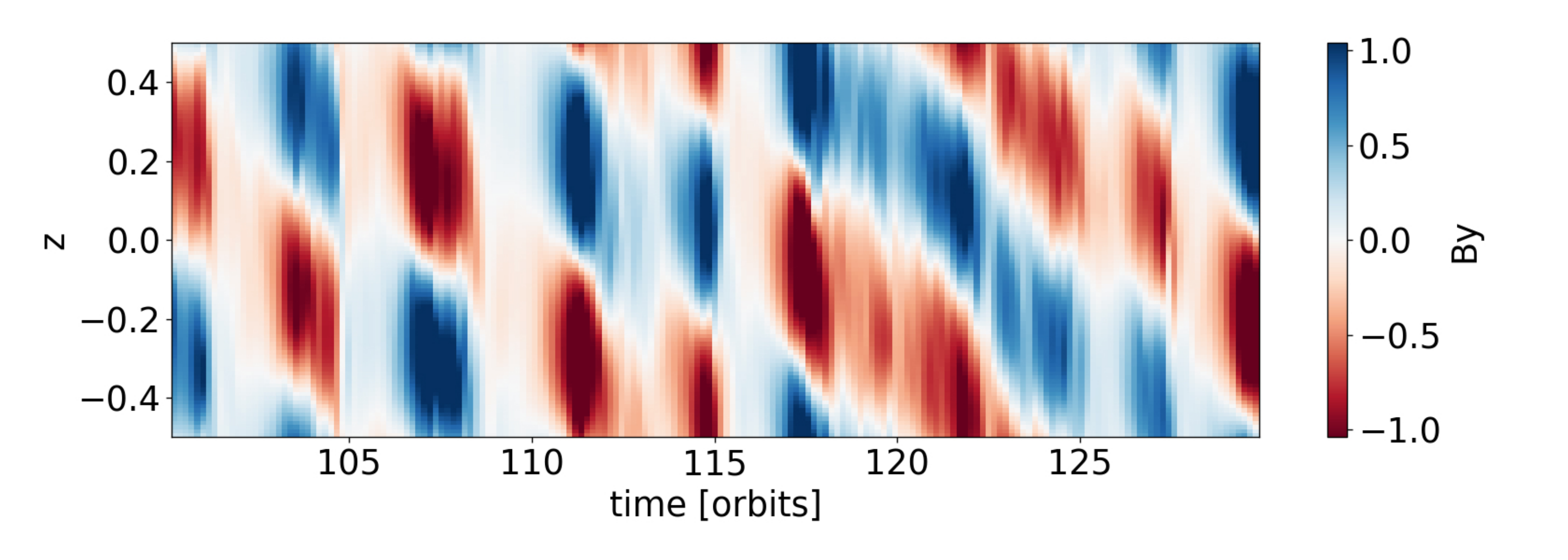}
    \includegraphics[width=9cm]{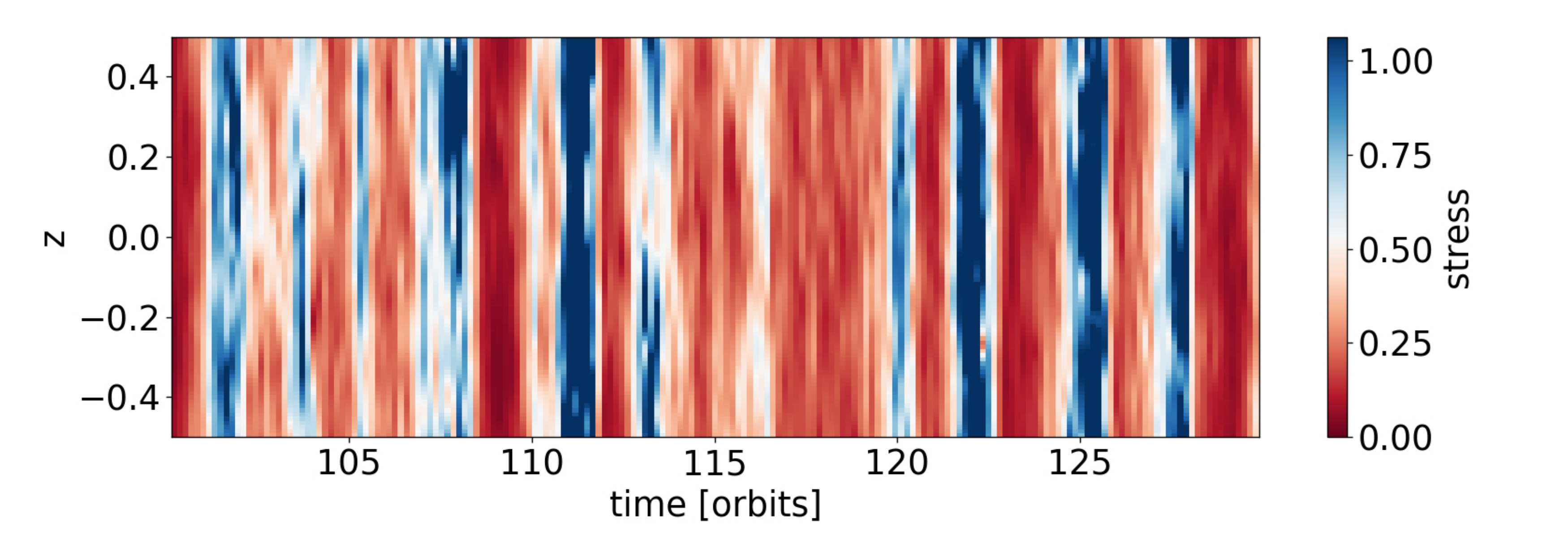}
    \includegraphics[width=9cm]{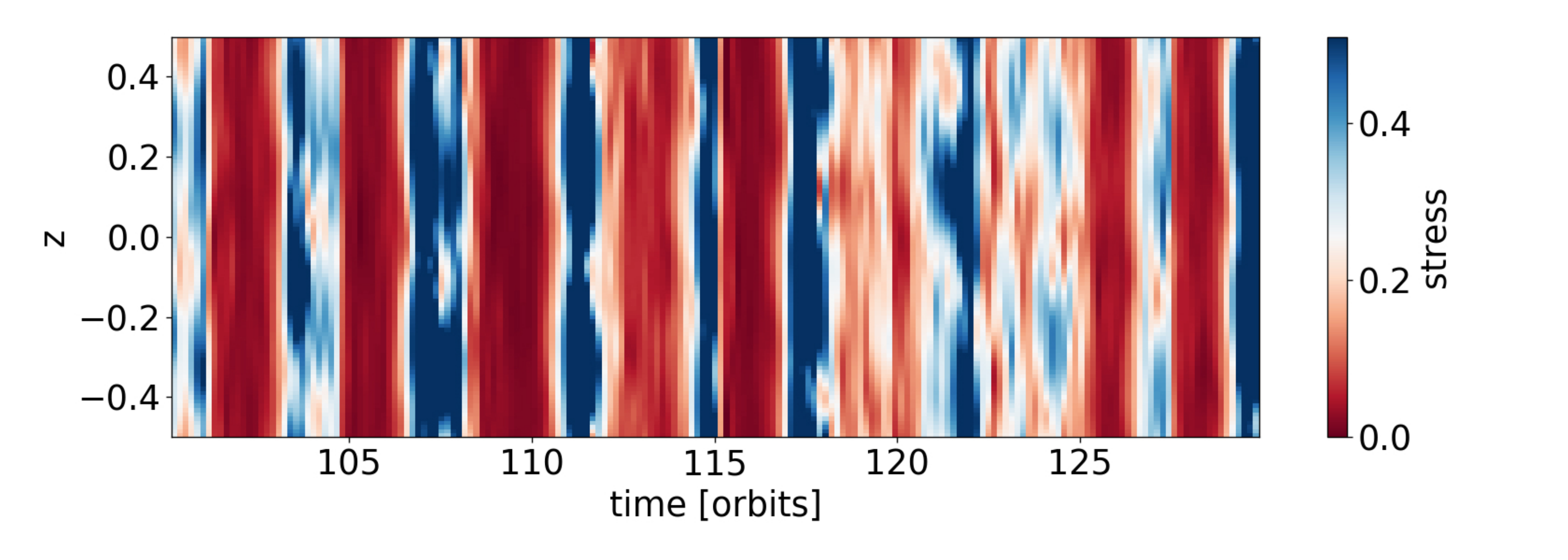}
    \caption{Spacetime diagrams showing the azimuthal magnetic field at the top and the total stress at the bottom. The left figures shows the simulation with an artificial resistivity coefficient $\alpha_{B}=0.5$ and the right figures show the simulation with $\alpha_{B}=4$. The figures clearly show the peaks related to the continuous creation and destruction of channel modes. Increasing the resistivity leads to a suppression of small scale magnetic fluctuation, and this means that more of the stress will be generated by the mean-field component.}
    \label{fig:UNFrender}
\end{figure*}
\begin{figure}[!h]
    \centering
    \includegraphics[width=\hsize]{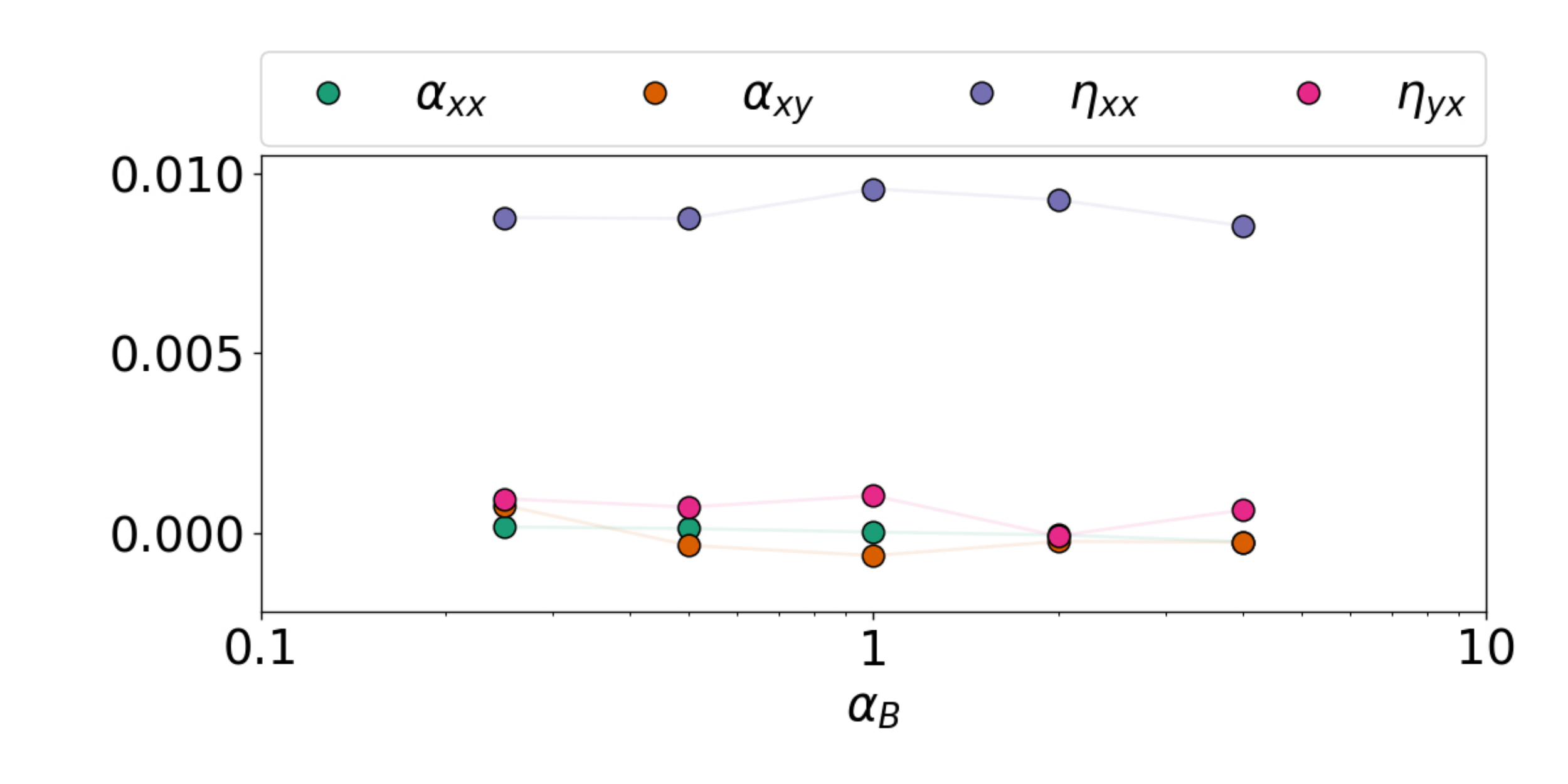}
    \caption{Time-averaged turbulent transport coefficient from the unstratified net-flux cases. To minimize noise/bias we have set $\alpha_{xx}=\alpha_{yy}$, $\eta_{xx}=\eta_{yy}$, $\alpha_{yx}=0.0$ and $\eta_{xy}=0.0$. We can see that only the turbulent resistivity $\eta_{xx}$ has a consistent value above $0.0$, with a value of about $0.008$. }
    \label{fig:UNFtransport}
\end{figure}

We setup our simulation with a shearing box of size $L=(1.0,\pi,1.0)$ with a resolution of $[n_x,n_y,n_z]=[48,150,48]$. The magnetic field is initialized with a constant vertical component
\begin{equation}
    B=\sqrt{\frac{2P_0}{\beta}}\hat{z},
\end{equation}
With a plasma beta of $\beta=400$ and pressure equal to $P_0=1.0$. Using \eq~\ref{eq:quality} we can see that we resolve $\lambda_{MRI}$ with a vertical quality parameter of $Q_z=[22,30,45]$ . We carry out several simulations with various artificial resistivity coefficients $\alpha_B=[0.25,0.5,1.0,2.0,4.0]$, where $\alpha_B=0.5$ is the code default. The simulations are run for about $200$ orbits or until the turbulence dies out. The results of the simulations are shown in \fig~\ref{fig:UNFtimeevo} to \ref{fig:UNFtransport}.
\\ \\
In \fig~\ref{fig:UNFtimeevo} we can see that the turbulent dynamo in all the simulations reach a saturated state with a heavily fluctuating magnetic energy density, which is what we expect from the unstratified NF case \citep{1995ApJ...440..742H} \footnote{We have also performed simulations with excessively strong dissipation $\Lambda<1$ and simulations with a very weak magnetic field (such that MRI is unresolved) to ensure that the MRI does not grow in these situations}.  From \fig~\ref{fig:UNFaverage}, we can see that as we decrease the resistivity, the magnetic energy and stress increases rapidly, where the total stress goes from $\alpha_{stress}=0.25$ to $0.9$. For the normalized magnetic stress($\alpha_{mag}$), we can see that the average lies around $0.65$ with only a weak dependency on the resistivity. The normalized magnetic stress is, in general, higher than what has been seen in previous Eulerian grid simulations, where $\alpha_{mag} \approx 0.4$ to $0.6$. The magnetic energy and Maxwell stress vary widely in the literature ($\alpha_{stress}=10^{-2}\rightarrow10^0$) and our values are similar to the ones reported in \citet{1995ApJ...440..742H} and \citet{2009ApJ...690..974S}. From \fig~\ref{fig:UNFaverage}, the ratio between the Maxwell stress and Reynolds stress($\alpha_{MW}/\alpha_{rey}$) shows a value of around $4.0$ with an increasing trend for lower resistivity. This is also similar to values reported in \citet{1995ApJ...440..742H} but somewhat lower than \citet{2009ApJ...690..974S} ($\alpha_{MW}/\alpha_{rey} \approx 7.6$). 
\\ \\
The higher $\alpha_{mag}$ can likely be explained by the use of a smaller box size $L=(1.0,\pi,1.0)$ compared to most other studies, which use $L=(1.0,2\pi,1.0)$. A smaller aspect ratio in the NF case does in general show stronger fluctuations in the turbulent state \citep{2008A&A...487....1B,2009MNRAS.396..779L}. The stronger fluctuations are a result of suppressing larger MRI modes that would otherwise participate in the non-linear dynamics, which heavily effect the growth and decay of channel modes. In \fig~\ref{fig:UNFchannelmode}, we can see an example of the formation and destruction of such a channel mode; during this process the magnetic energy and stress will peak. In addition to being dependent on the aspect ratio of the box, the growth and destruction of these channel modes will depend on the dissipation. This can clearly be seen in \fig~\ref{fig:UNFrender} where we see the evolution of the horizontal averaged azimuthal field and stress over a period of thirty orbits for two different resistivities($\alpha_{B}=0.5$ and $\alpha_{B}=4$). We can see that during this time channel modes are subsequently formed and destroyed, but with different frequency and behavior. In the high resistivity case, the magnetic energy peaks during channel mode formation but most of the small-scale magnetic fluctuations are quickly suppressed after channel mode breakdown. This means that less of the stress within this case comes from the turbulent component. This can also be seen in \fig~\ref{fig:UNFaverage}, where the mean-field component dominates over the turbulent component at $\alpha_{B}=4$ while becoming almost equal at $\alpha_{B}=0.5$. Interestingly, the normalized mean-field and turbulent magnetic stresses stays fairly constant $\alpha_{mag,mean}\approx0.8$ and $\alpha_{mag,turb} \approx 0.55$.
\\ \\
The average divergence error remains either below or close to $\epsilon_{div,err}\approx 10^{-2}$. For the simulations with $\alpha_B=[0.25,0.5,1.0,2.0,4.0]$ the corresponding time-averaged Prandtl numbers is $\Vmt{P_m} =[1.95,1.42,0.96,0.60,0.35]$. The standard default value of $\alpha_B=0.5$ has a $P_m\approx1.5$. The Elsasser number remains far above $1$ for all the cases and the average plasma beta rises linearly with $\alpha_B$ with value between $\beta \approx 5 \rightarrow 20$ 
\\ \\
\fig~\ref{fig:UNFtransport} shows the time-averaged values of the transport coefficients $\alpha_{xx}$, $\alpha_{xy}$, $\eta_{xx}$ and $\eta_{yx}$ for all the simulations. From the figure, we can see that both $\alpha$ coefficients have values very close to zero which is to be expected from the unstratified case. $\eta_{yx}$ does also not have a significant value and remains close to zero. The only value that has a significant value above zero is the turbulent diffusivity which has a value of around $\eta_{xx}\approx0.008$.
\subsection{Zero net-flux simulations}
\label{subsec:ZNF}
\begin{figure*}[!h]
    \centering
    \includegraphics[width=16cm]{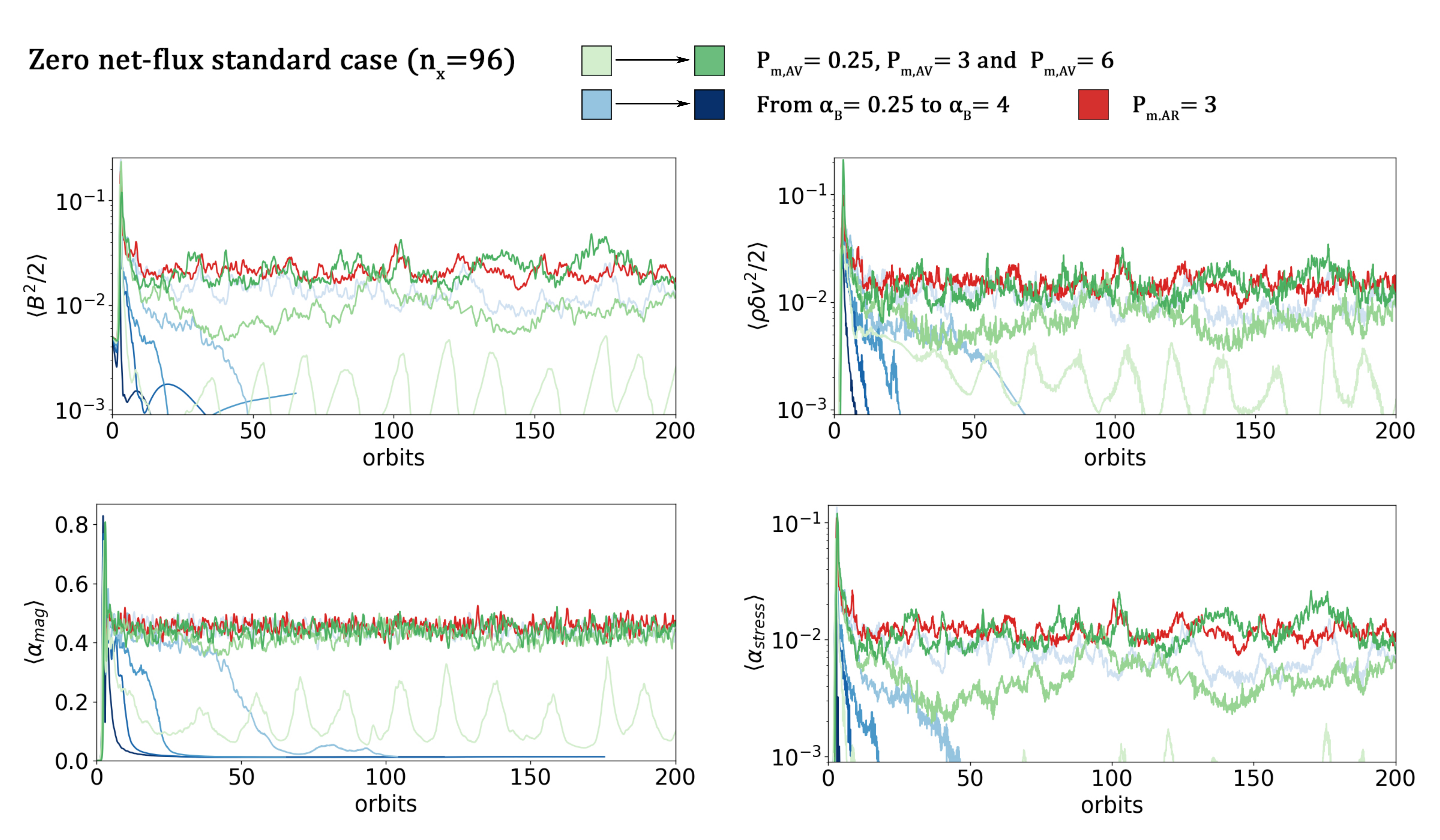}
    \caption{Time evolution of several volume-averaged quantities for the standard box, unstratified ZNF cases at a resolution of $n_x=96$: magnetic energy(top left), kinetic energy(top right), normalized Maxwell stress(bottom left) and the total stress(bottom right). The red line show the case of $P_{m,AR}=3.0$ where we set the Prandtl number by altering the AR strength. The green lines show the case where we set the Prandtl number by altering the AV strength, the darkness of the curve is determined by the value of the set Prandtl number, $P_{m,AV}=[0.25, 3.0 ,6.0]$. The blue curves represent the cases with a set AR coefficient without forcing the Prandtl number, where the darkness is determined by the strength of the artificial resistivity parameter, $\alpha_B=[0.25,0.5,1.0,2.0,4.0]$. Four of the nine cases reach a saturated state ($\alpha_B=0.25$, $P_{m,AR}=3.0$, $P_{m,AV}=3.0$, $P_{m,AV}=6.0$) The code default value of $\alpha_B=0.5$ sustains turbulence for around 50 orbits before decaying.}
    \label{fig:UZNFtimeevo}
\end{figure*}
\begin{figure*}[!h]
    \centering
    \includegraphics[width=16cm]{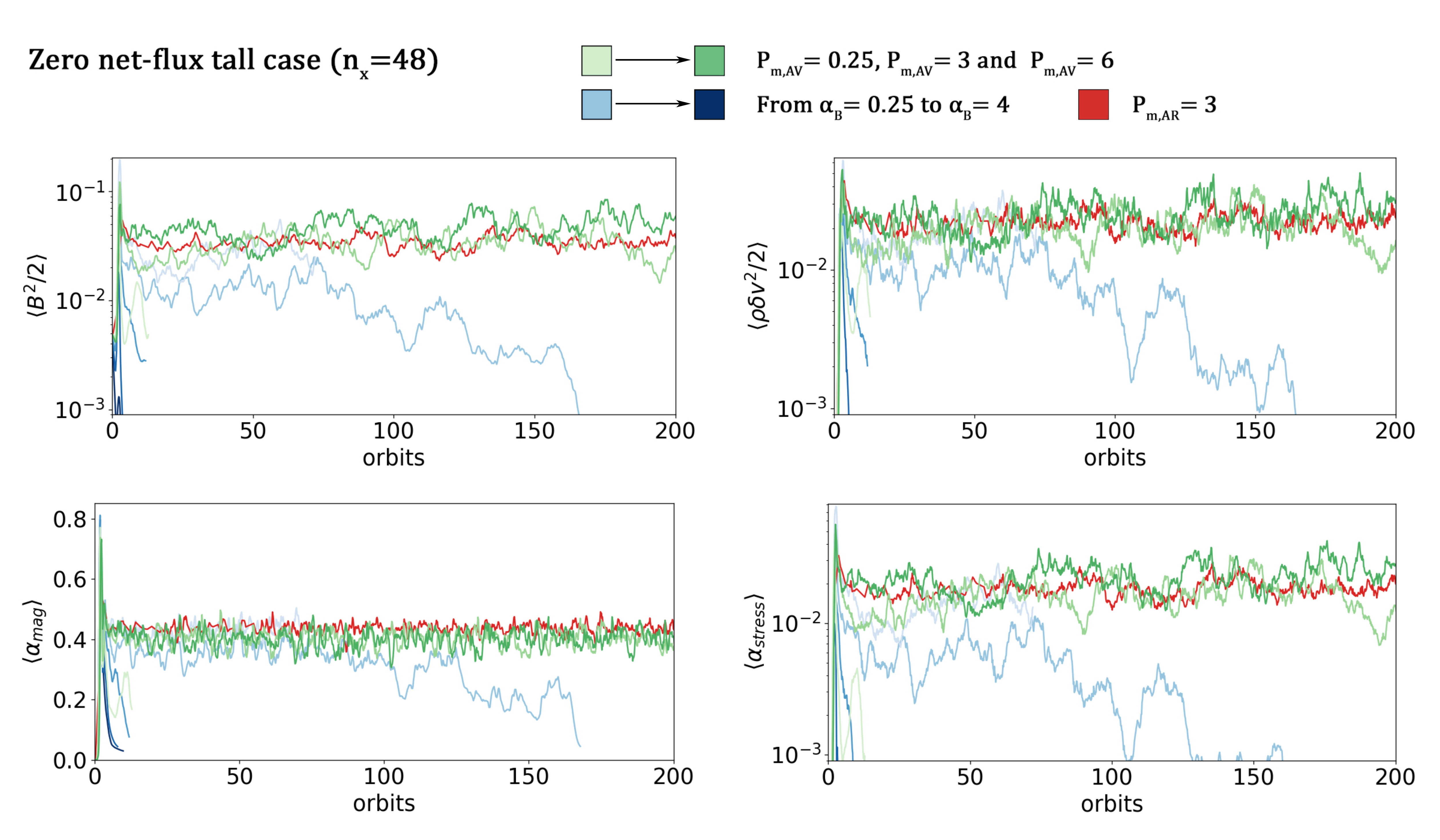}
    \caption{Time evolution of several volume-averaged quantities for the tall box, unstratified ZNF cases at a resolution of $n_x=48$: magnetic energy(top left), kinetic energy(top right), normalized Maxwell stress(bottom left) and the total stress(bottom right). The red line show the case of $P_{m,AR}=3.0$ where we set the Prandtl number by altering the AR strength. The green lines show the case where we set the Prandtl number by altering the AV strength, the darkness of the curve is determined by the value of the set Prandtl number, $P_{m,AV}=[0.25, 3.0 ,6.0]$. The blue curves represent the cases with a set AR coefficient without forcing the Prandtl number, where the darkness is determined by the strength of the artificial resistivity parameter, $\alpha_B=[0.25,0.5,1.0,2.0,4.0]$. Four of the nine cases reach a saturated state ($\alpha_B=0.25$, $P_{m,AR}=3.0$, $P_{m,AV}=3.0$, $P_{m,AV}=6.0$).The code default value of $\alpha_B=0.5$ sustains turbulence for a long time but starts to decay after around 120 orbits.}
    \label{fig:UZNFTtimeevo}
\end{figure*}
\begin{figure*}[!h]
    \centering
    \includegraphics[width=14cm]{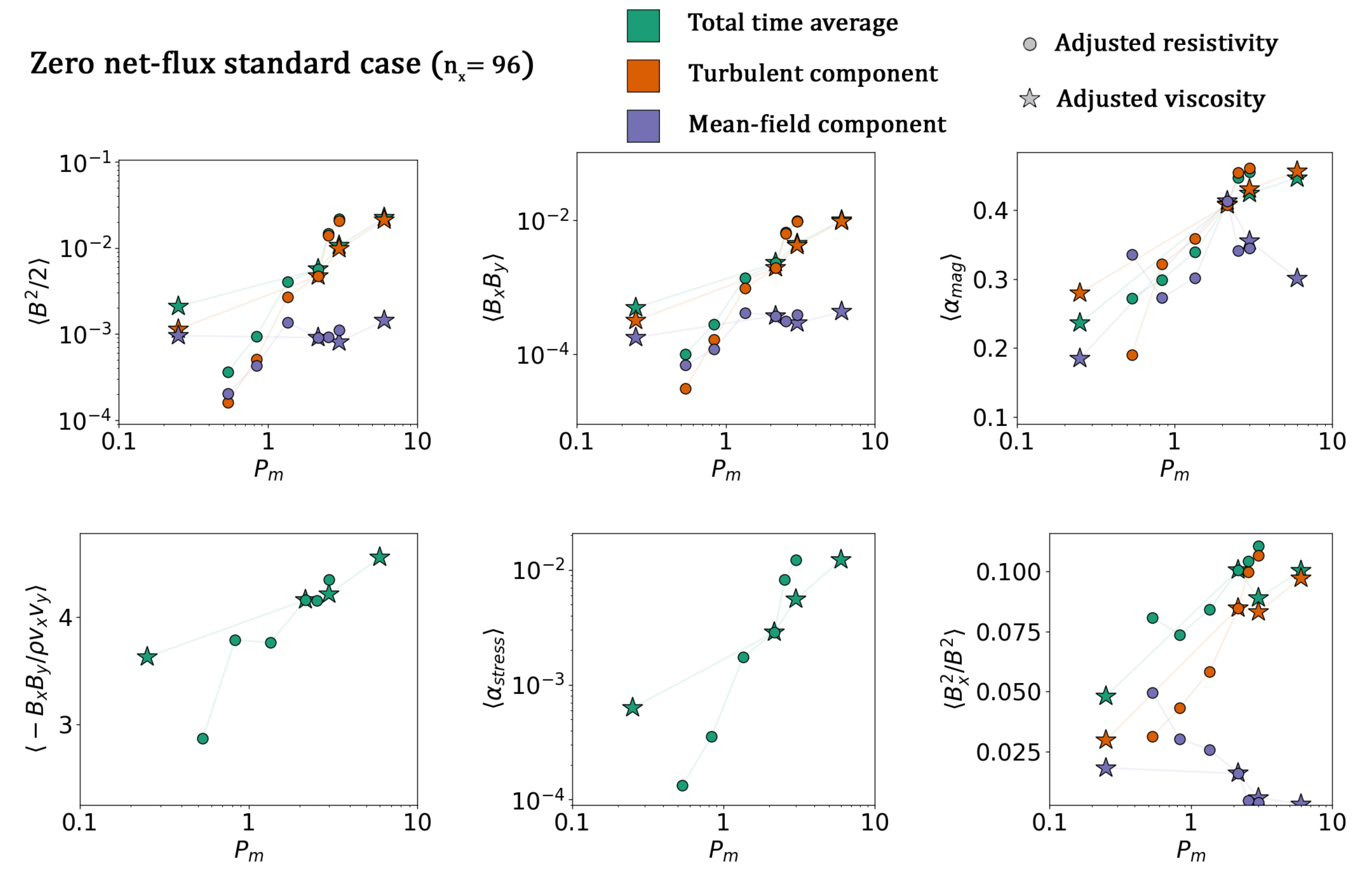}
    \caption{Time-averaged values of several quantities for all our unstratified, zero net-flux simulations with standard box size ($L=[1.0,\pi,4.0]$) and resolution $n_x=96$. From the top left to bottom right, we plot the magnetic energy density, the Maxwell stress, the normalized Maxwell stress, the ratio between Reynolds and Maxwell stresses, the total stress, the ratio between radial and total magnetic field energy. The x-axis shows the time averaged effective Prandtl number of the simulation. For some quantities we have plotted the total time average(shown in green), the time average of the turbulent component(shown in orange) and the time average of the mean component(shown in blue). The circles represent the simulations where we have adjusted the strength of the artificial resistivity, while the star symbols represent the simulations where we have adjusted the artificial viscosity. }
    \label{fig:UZNFaverage}
\end{figure*}
\begin{figure*}[!h]
    \centering
    \includegraphics[width=14cm]{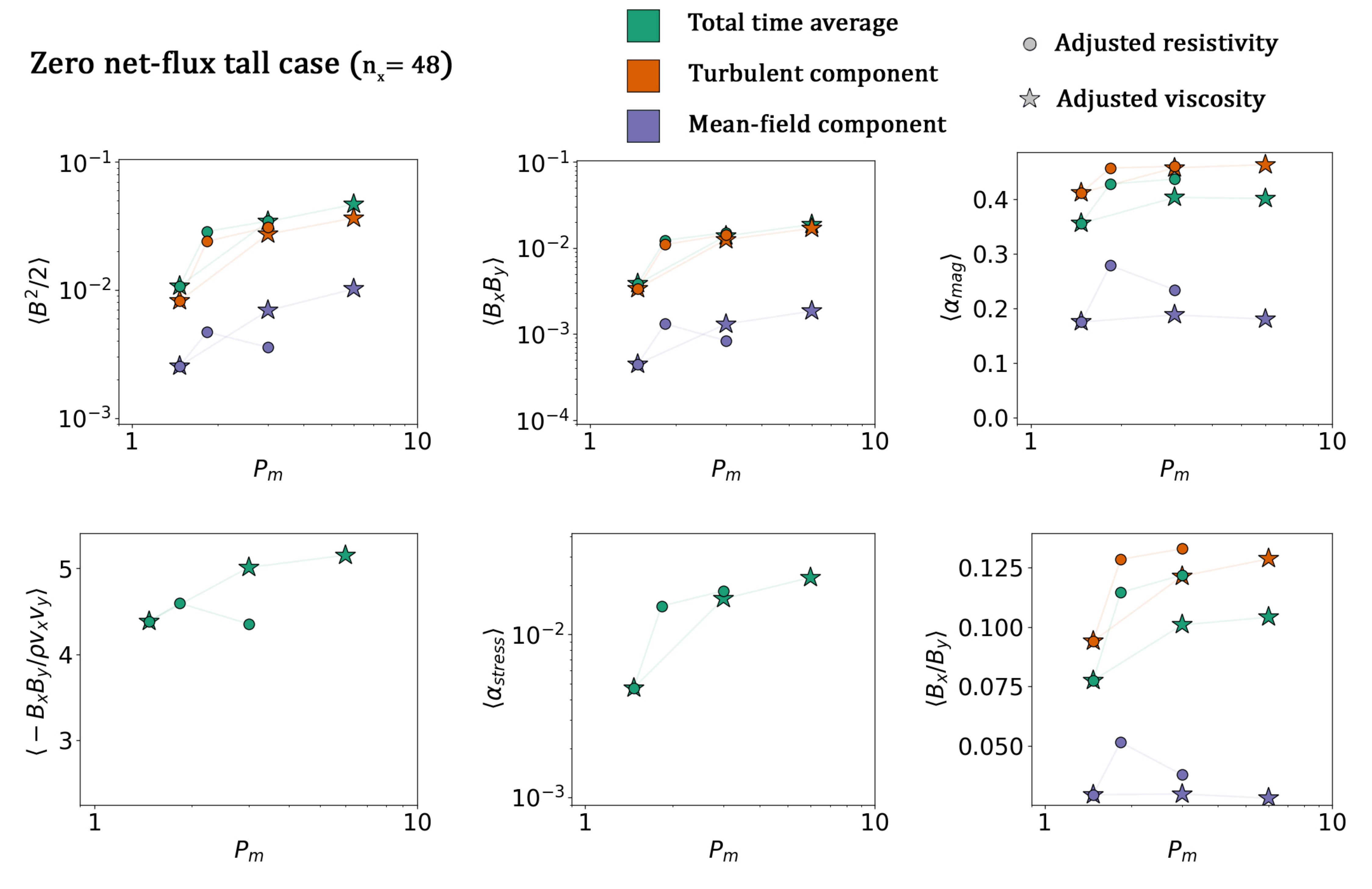}
    \caption{Time-averaged values of several quantities for all our unstratified, zero net-flux simulations with tall box size ($L=[1.0,\pi,4.0]$) and resolution $n_x=48$. From the top left to bottom right, we plot the magnetic energy density, the Maxwell stress, the normalized Maxwell stress, the ratio between Reynolds and Maxwell stresses, the total stress, the ratio between radial and total magnetic field energy. The x-axis shows the time averaged effective Prandtl number of the simulation. For some quantities we have plotted the total time average(shown in green), the time average of the turbulent component(shown in orange) and the time average of the mean component(shown in blue). The circles represent the simulations where we have adjusted the strength of the artificial resistivity, while the star symbols represent the simulations where we have adjusted the artificial viscosity.}
    \label{fig:UZNFTaverage}
\end{figure*}
\begin{figure*}[!h]
\centering
    Standard ZNF ($P_{m,AV}=3$) \qquad \qquad \qquad \qquad \qquad \qquad \qquad \qquad \qquad  Tall ZNF ($P_{m,AV}=3$)
    \includegraphics[width=9cm]{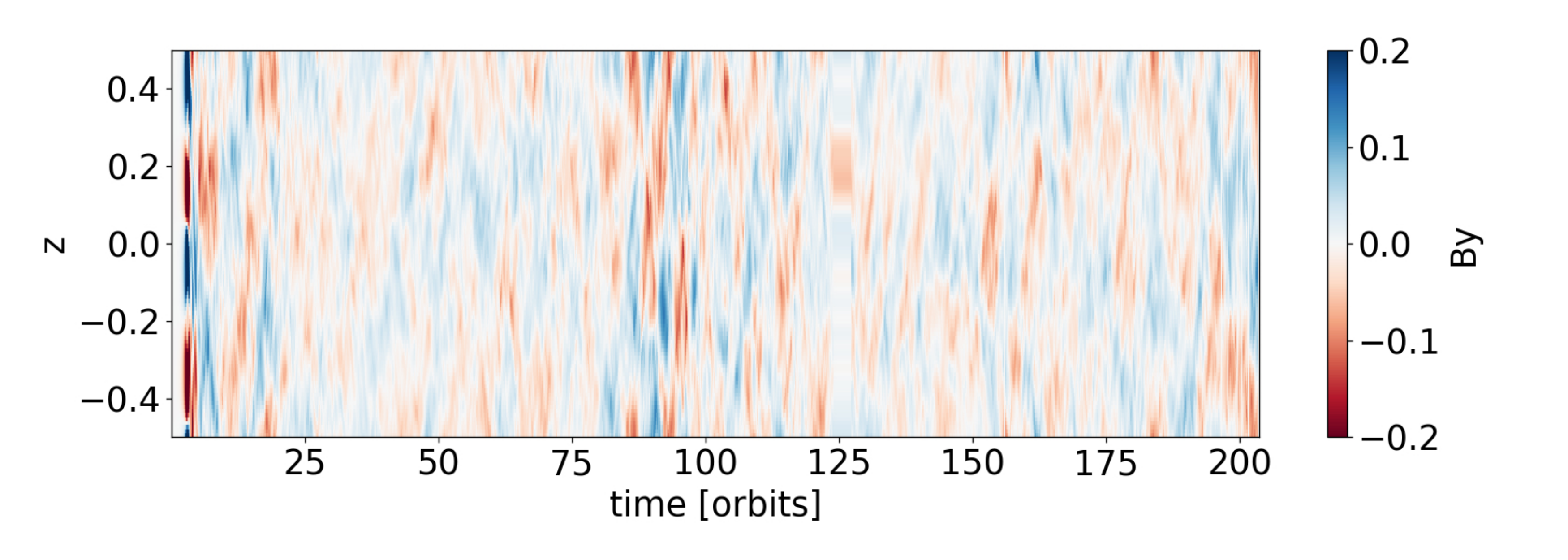}
    \includegraphics[width=9cm]{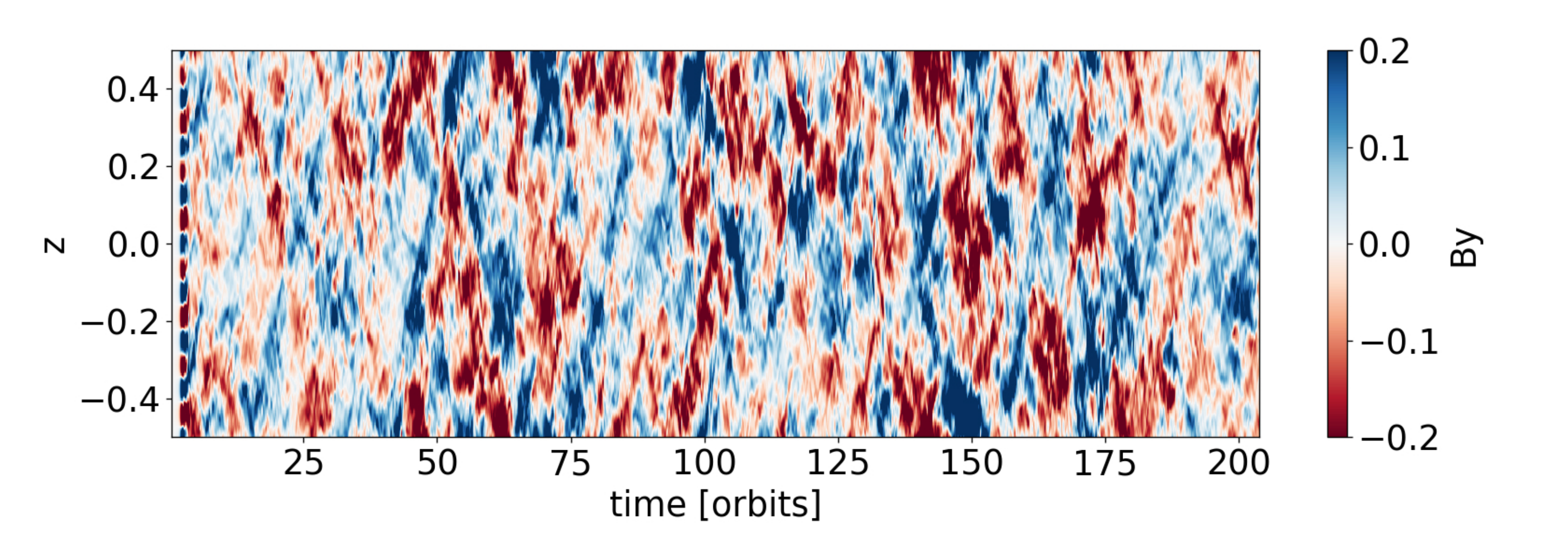}
    \caption{Spacetime diagrams showing the horizontal-averaged azimuthal magnetic field for the unstratified, zero net flux case ($P_{m,AV}=3$). We can see that the tall box has larger structured mean-fields than the standard box. Compared to \cite{2016MNRAS.456.2273S} the rendered pattern of the mean-fields are similar, however, they produce much stronger mean-fields in their tall box simulations.}
    \label{fig:UZNFrender}
\end{figure*}
\begin{figure*}[!h]
    \centering
    \includegraphics[width=\hsize]{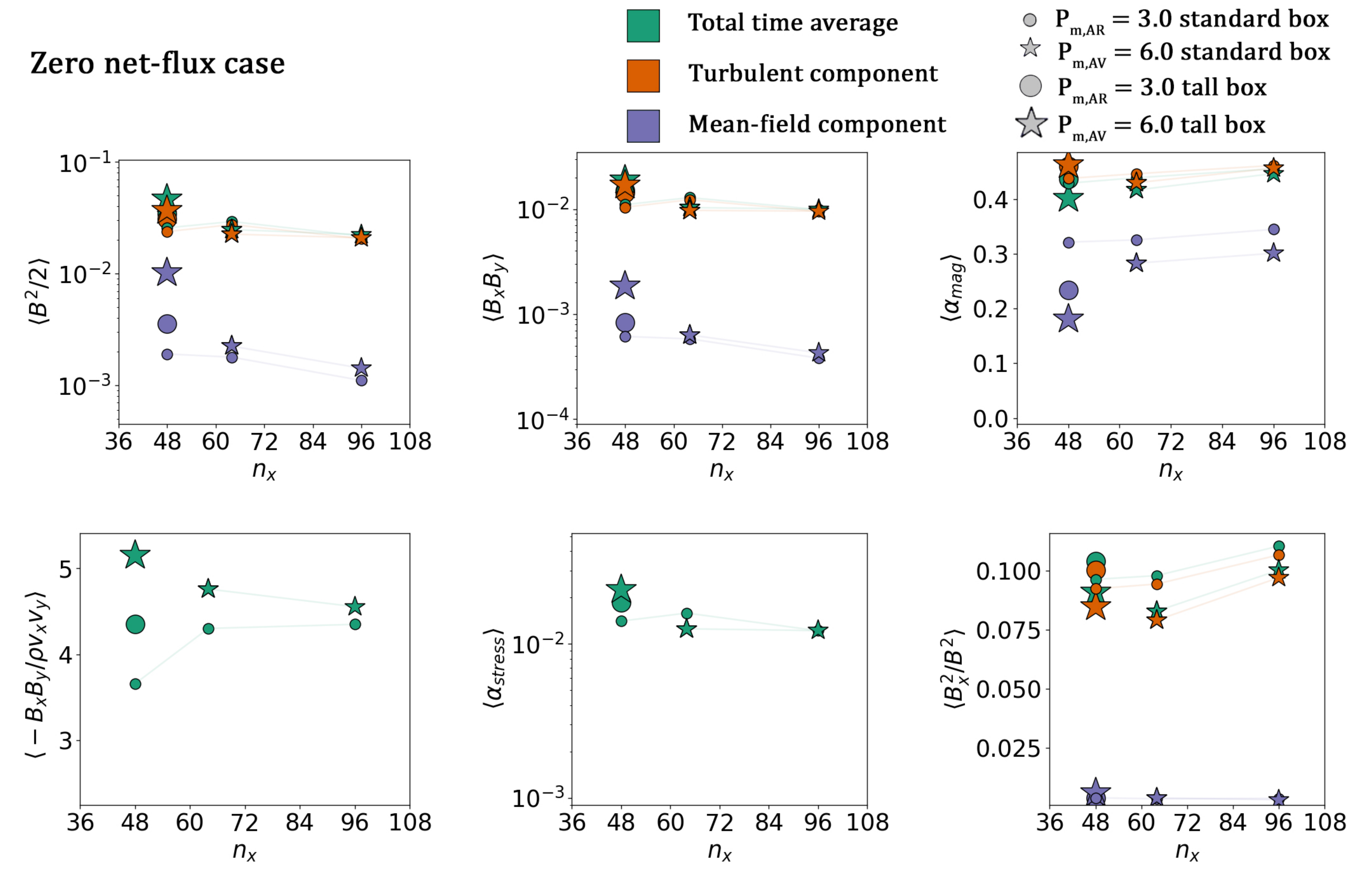}
    \caption{Time-averaged values of several quantities for our resolution study of the unstratified, zero net-flux  cases with $P_{m,AR}=3.0$ and $P_{m,AV}=6.0$, which includes tall box and standard box simulations. From the top left to bottom right, we plot the magnetic energy density, the Maxwell stress, the normalized Maxwell stress, the ratio between Reynolds and Maxwell stresses, the total stress, the ratio between radial and total magnetic field energy. For some quantities we have plotted the total time average(shown in green), the time average of the turbulent component(shown in orange) and the time average of the mean component(shown in blue). The circles represent the simulations where we have adjusted the strength of the artificial resistivity, small circles represent standard box cases and large circles represent tall box. While the star (standard box size) and large stars (tall box size) represent the simulations where we have adjusted the artificial viscosity.}
    \label{fig:UZNFresolution}
\end{figure*}
\begin{figure}[!h]
    \centering
    \includegraphics[width=\hsize]{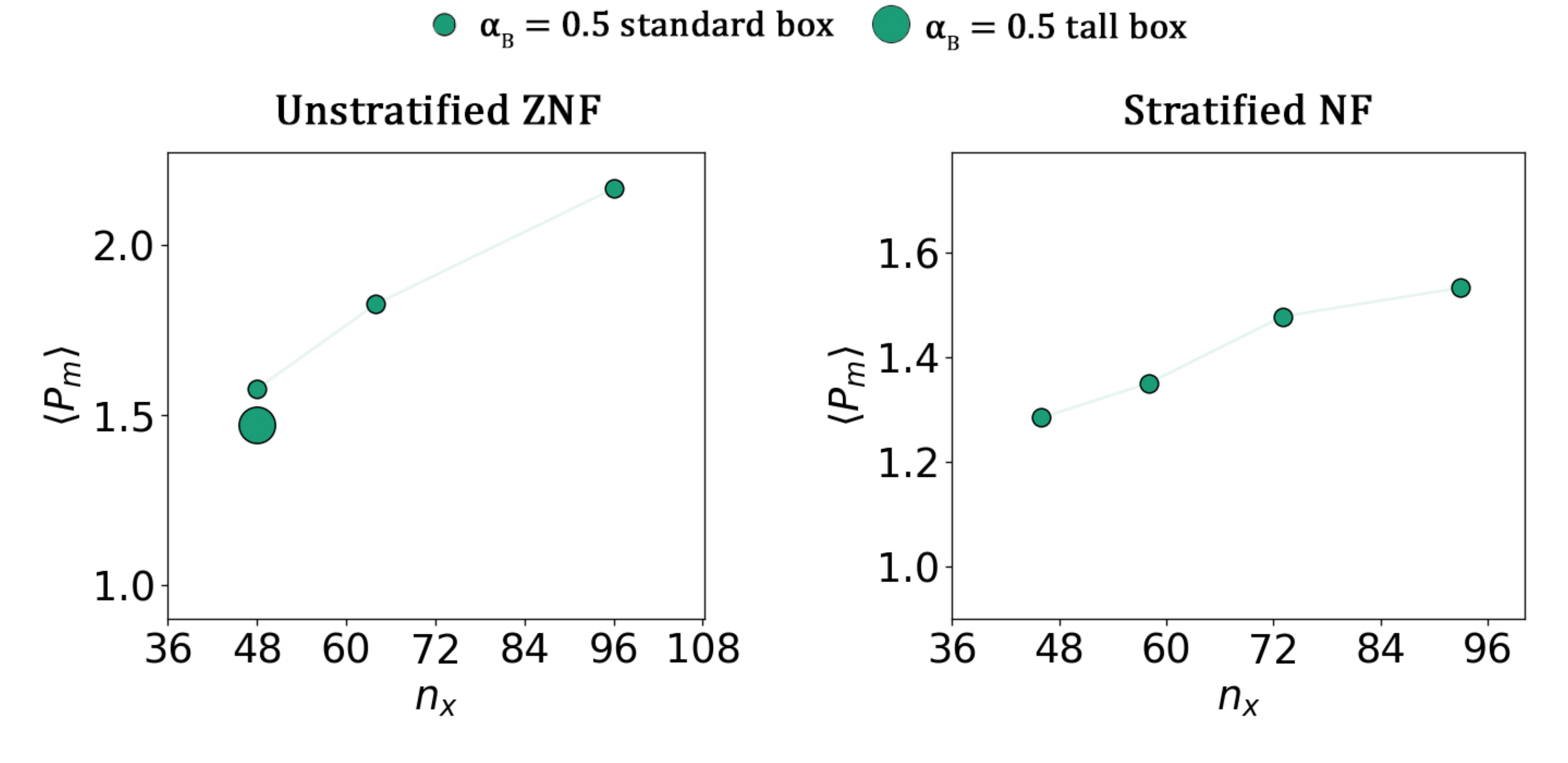}
    \caption{Resolution dependence of the numerical Prandtl number for the unstratified, zero-flux cases on the left (section \ref{subsec:ZNF}) and for the stratified, net-flux cases to the right (section \ref{sec:SNF}). This shows cases with an AR coefficient set to $\alpha_B=0.5$, which is our code default. We can see that we have an almost linear increase with higher resolution for both cases.}
    \label{fig:Pmresdepend}
\end{figure}
\begin{figure*}[!h]
    \includegraphics[width=\hsize]{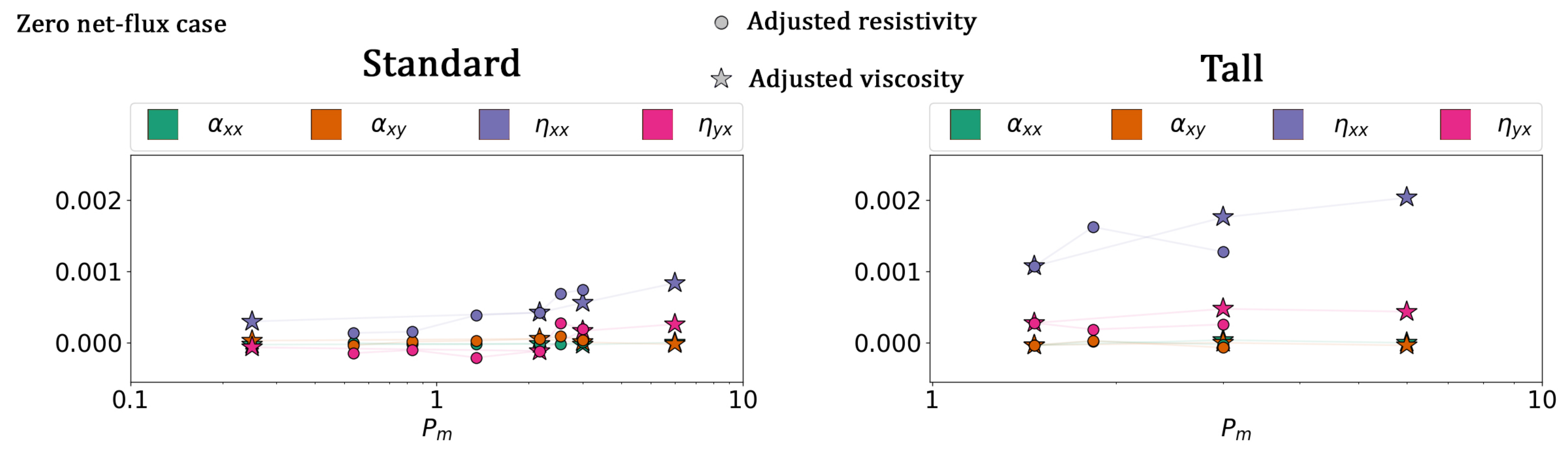}
    \caption{Time-averaged turbulent transport coefficients from the unstratified, zero net-flux cases, with the standard box size cases ($L=[1.0,\pi,1.0]$, $n_x=96$) to the left and the tall box cases ($L=[1.0,\pi,4.0]$, $n_x=48$) to the right. The circles represent the simulations where we have adjusted the strength of the artificial resistivity, while the star symbols represent the simulations where we have adjusted the artificial viscosity. To minimize noise we have set $\alpha_{xx}=\alpha_{yy}$, $\eta_{xx}=\eta_{yy}$, $\alpha_{yx}=0.0$ and $\eta_{xy}=0.0$. We can see that the $\alpha$ coefficients have values of around zero as expected for the unstratified case and $\eta_{xx}$ with a consistent positive value. For the standard box $\eta_{yx}$ remain close to zero while for the tall box we see a consistent positive value which is contrary to what was seen by \cite{2016MNRAS.456.2273S}.}
    \label{fig:UZNFtransport}
\end{figure*}

The setup follows from \cite{2019ApJS..241...26D}, in which a shearing box together with a varying vertical magnetic field is initialized
\begin{equation}
    B=B_0\hat{z}\sin(2\pi x).
\end{equation}{}
Here, $B_0$ is the initial magnetic field strength and is set such that the volume averaged plasma beta is $\beta=2P/B^2=400$. We run the simulations at three different resolutions $[n_x,n_y,n_z]=[48,150,48], [64,200,64], [96,300,96]$ with a standard box with length $L=(1.0,\pi,1.0)$. To test the effect of a taller box within SPH, we also run with a domain size of $L=(1.0,\pi,4.0)$ (same as in \cite{2016MNRAS.456.2273S}), with a resolution of $[n_x,n_y,n_z]=[48,150,192]$. Using \eq~\ref{eq:quality}, we can see that we resolve $\lambda_{MRI}$ with an initial quality parameter of $Q_z=[22,30,45]$ in each respective resolution. We carry out several simulations at each resolution by varying the artificial resistivity coefficient $\alpha_B=[0.25,0.5,1.0,2.0,4.0]$ (blue curves in \fig~\ref{fig:UZNFtimeevo} and \fig~\ref{fig:UZNFTtimeevo}), where $\alpha_B=0.5$ is the default value. The corresponding time-averaged Prandtl numbers in the $n_x=96$ standard box is $\Vmt{P_m} =[2.54,2.17,1.35,0.83,0.54]$ and in the $n_x=48$ tall box $\Vmt{P_m} =[1.84,1.47,1.11,0.67,0.34]$.  In addition, we run four cases where we force a certain average numerical Prandtl number by adjusting either the artificial viscosity or the artifical resistivity (see \sect~\ref{subsec:simsetup}). One case is run by adjusting the artificial resistivity, where we force the Prandtl number to be equal to $P_{m,AR}=3$ (the red curve in \fig~\ref{fig:UZNFtimeevo} and \fig~\ref{fig:UZNFTtimeevo}). Two of the cases adjust the artificial viscosity with $\alpha_B=0.5$, forcing a Prandtl number of $P_{m,AV}=3$ and $P_{m,AV}=6$ and one case adjust the artificial viscosity with $\alpha_B=0.25$, forcing a Prandtl number of $P_{m,AV}=0.25$ (green curves in \fig~\ref{fig:UZNFtimeevo} and \fig~\ref{fig:UZNFTtimeevo}). The simulations are run for about $200$ orbits or until the turbulence dies out. The results of the simulations are shown in \fig~\ref{fig:UZNFtimeevo} to \ref{fig:UZNFtransport}.
\\ \\
In \fig~\ref{fig:UZNFtimeevo} we show the time evolution of the magnetic energy, kinetic energy, normalized Maxwell stress, and the total stress for our high-resolution standard box cases with $n_x=96$. Only four of the nine cases reach a saturated state ($\alpha_{B}=0.25$, $P_{m,AR}=3$, $P_{m,AV}=3$, $P_{m,AV}=6$), which all have a Prandtl number of $P_m>2.5$. The $\alpha_{B}=0.5$ case sustain turbulence for about 50 orbits before decaying, and most of the other cases have their turbulence eliminated within the first 30 orbits, similar to what was seen in \cite{2019ApJS..241...26D}, where the longest living case was about 20 orbits. An outlier is the evolution of $P_{m,AV}=0.25$, where small stress oscillation can still be seen for a long time after the initial decay. This is simply caused by numerical noise, as we force a very low AV for this case together with a low AR coefficient $\alpha_B=0.25$. In \fig~\ref{fig:UZNFTtimeevo}, we show the time evolution of the same quantities for the tall-box simulations with resolution $n_x=48$. We find that the same four cases reach a saturated state for the tall box ($\alpha_B=0.25$,$P_{m, AR}=3$,$P_{m,AV}=3$,$P_{m,AV}=6$). The $\alpha_B=0.5$ case sustain turbulence for a longer time, decaying after around 120 orbits. Most of the other cases have their turbulence eliminated within the first 20 orbits.
\\ \\
In \fig~\ref{fig:UZNFaverage} and \fig~\ref{fig:UZNFTaverage} we show the time-averaged quantities of the high-resolution($n_x=96$), standard box runs and the lower resolution ($n_x=48$), tall box cases, respectively. For the tall box case, we only show the time averages of the saturated runs as all the other runs are killed after their initial turbulence die out. For the standard box, we can see that as we increase the Prandtl number the magnetic energy and stress increases rapidly until we reach a high enough $P_m$ for saturation. The saturated cases reaches a total stress of around $\alpha_{stress}=0.01$ and a normalized magnetic stress of $\alpha_{mag}=0.4$, which is consistent with previous studies of the MRI \citep{1995ApJ...440..742H,2009ApJ...690..974S}. From these figures, we can see that the total magnetic field energy and stresses are largely dominated by the turbulent component, with only a very weak mean-field component. The mean-field energy and stress do not change significantly as the Prandtl number is increased. We can also see that there is not a one to one correlation between energies and stresses for $P_{m,AR}$ and $P_{m,AV}$ simulations with the same Prandtl number. The resulting stress levels are higher when lowering the artificial resistivity compared to increasing the artificial viscosity. While saturation is mainly governed by the Prandtl number, stress levels will depend on both the Prandtl number and the strength of the resistivity \citep{2009ApJ...707..833S}. In addition, as shown in \cite{2007A&A...476.1123F} the critical Prandtl number does have a dependence on the Reynolds number which also can have an effect on the amplitude of the saturated stress.
\\ \\
Comparing this to the time average results from the tall box shown in \fig~\ref{fig:UZNFTaverage}, we can see that similar to the standard box, magnetic energy and stress increases with $P_m$, reaching albeit higher values but with similar normalized magnetic stress values. From the magnetic energy density, it is clear that the turbulent part of the energy is dominant similar to the standard box case. This is quite different from the results presented in \cite{2016MNRAS.456.2273S} where the mean-field contributes most to the energy density. In fact, they showed a rapid increase in mean-field energy density as the vertical aspect ratio of the domain was increased. We do see that the mean-field energy density of our tall box is larger than the standard box, however, it remains relatively weak. A visual comparison of the mean-fields can be seen in \fig~\ref{fig:UZNFrender}, where we see that the tall box has larger structured mean-fields than the standard box. The rendered pattern of the mean-fields are reminiscent of the result presented in \cite{2016MNRAS.456.2273S}. The lack of significant mean-fields can be seen in the resulting stress levels of our simulations, which are only slightly larger than the ones from the standard box case and nowhere near the $\alpha_{stress}=10^{-1}$ presented in \cite{2016MNRAS.456.2273S}.
\\ \\
We can see from \fig~\ref{fig:UZNFaverage} and \fig~\ref{fig:UZNFTaverage} that, in general, there is a small increase in the ratio between Maxwell stress and Reynolds stress as Prandtl number is increased and seem to converge towards a value of roughly $\alpha_{MW} / \alpha_{rey} \approx 4.5$ for the standard box and $\alpha_{MW} / \alpha_{rey} \approx 5$ in the tall box. For the standard box, this is slightly higher than the typical values from Eulerian grid simulations, which report values of around $\alpha_{MW}/\alpha_{rey}\approx 3 \leftrightarrow 4$ \citep{1995ApJ...440..742H,1996MNRAS.281L..21A,1996ApJ...463..656S,1999ApJ...518..394H,2004ApJ...605..321S}. However the tall box values are in accordance with those presented in \cite{2016MNRAS.456.2273S}.
\\ \\
The average divergence error remains either below or close to $\epsilon_{div,err}\approx 10^{-2}$. As expected, the divergence error is kept lower by increasing the artificial viscosity to reach a certain Prandtl number than by decreasing the resistivity. For the majority of cases, the Elsasser number remains far above $1$, however, for the high resistivity cases ($\alpha_B=4, \alpha_B=2$) the number drops below one, which is likely why we see such a rapid decay of turbulence in these cases. We also show the relative radial energy ratio ($B_x^2/B^2$), which shows a steep increase with $P_{m}$ and for our saturated runs it reaches values around $B_x^2/B^2 \approx 0.1$. This is similar to the values reported by \cite{1996ApJ...464..690H} but is somewhat lower than the higher resolution simulation performed by \cite{2009ApJ...690..974S} and \cite{2016MNRAS.456.2273S}, which reports values of around $B_x^2/B^2 \approx 0.14$. Looking at \fig~\ref{fig:UZNFresolution}, this is consistent with the increasing trend with resolution that we see. Increasing the vertical domain size slightly increases the value from about $B_x^2/B^2\approx0.09$ for the standard box to around $B_x^2/B^2\approx0.11$ for the tall box. This is opposite to what is found in \cite{2016MNRAS.456.2273S}, which see a consistent decrease in this value as the vertical domain size is increased, going from $B_x^2/B^2\approx0.14$ for the standard box to $B_x^2/B^2\approx0.12$ for the 4 times vertical ratio and $B_x^2/B^2\approx0.09$ for the 8 times vertical ratio.
\\ \\
We also performed a resolution study on the standard box case to see how different time-averaged quantities change with resolution. We primarily look at the two cases where we set $P_{m,AV}=6$ and $P_{m,AR}=3$. From \fig~\ref{fig:UZNFresolution}, we can see that for the saturated cases there is no strong resolution dependence on the total stress as reported by studies using Eulerian grid codes\citep{2006A&A...457..371F}. Instead, the stress saturates at around $\alpha_{stress}=0.01$. This resolution independence is of course only for the cases where we force a certain numerical Prandtl number, as increasing the resolution for a fixed resistivity coefficient will alter the numerical Prandtl number. The resolution dependency of the numerical Prandtl number can be seen in \fig~\ref{fig:Pmresdepend}, which shows that $P_{m}$ has an almost linear increase with resolution.
The normalized turbulent stress ratio remains fairly constant at around $\alpha_{mag} \approx 0.42$ with a slight increase with resolution. The relative radial energy ratio ($B_x^2/B^2$) show a steady increase with resolution and have not converged for our highest resolution case. The divergence error is also reduced with increasing resolution, which is consistent with our cleaning scheme implementation.
\\ \\
\fig~\ref{fig:UZNFtransport} shows the time averaged values of the transport coefficients $\alpha_{xx}$, $\alpha_{xy}$, $\eta_{xx}$ and $\eta_{yx}$ for all the unstratified simulations. From the figures, we can see that both $\alpha$ coefficients have values very close to zero which is to be expected from the unstratified case. $\eta_{yx}$ can be seen to have a slightly positive value for all cases exhibiting sustained turbulence, which is more significant in the tall box. The turbulent diffusivity can be seen to have a consistent positive value, which is around $\eta_{xx}\approx0.006$ for the standard box and $\eta_{xx}\approx0.015$ in the tall box. The non-negative value in $\eta_{yx}$ might explain why we do not see the generation of large mean-fields within our simulations as \cite{2016MNRAS.456.2273S} shows a consistent negative value for $\eta_{yx}$ which as we explained in the introduction can act to generate local mean-fields through the shear-current effect. However, the lack of shear-current effect is consistent with other previous studies of the MRI \citep{1995ApJ...446..741B,2008AN....329..725B,2010MNRAS.405...41G}

\section{Stratified simulation results}
\label{sec:SNF}
\begin{figure*}[!h]
    \centering
    \includegraphics[width=\hsize]{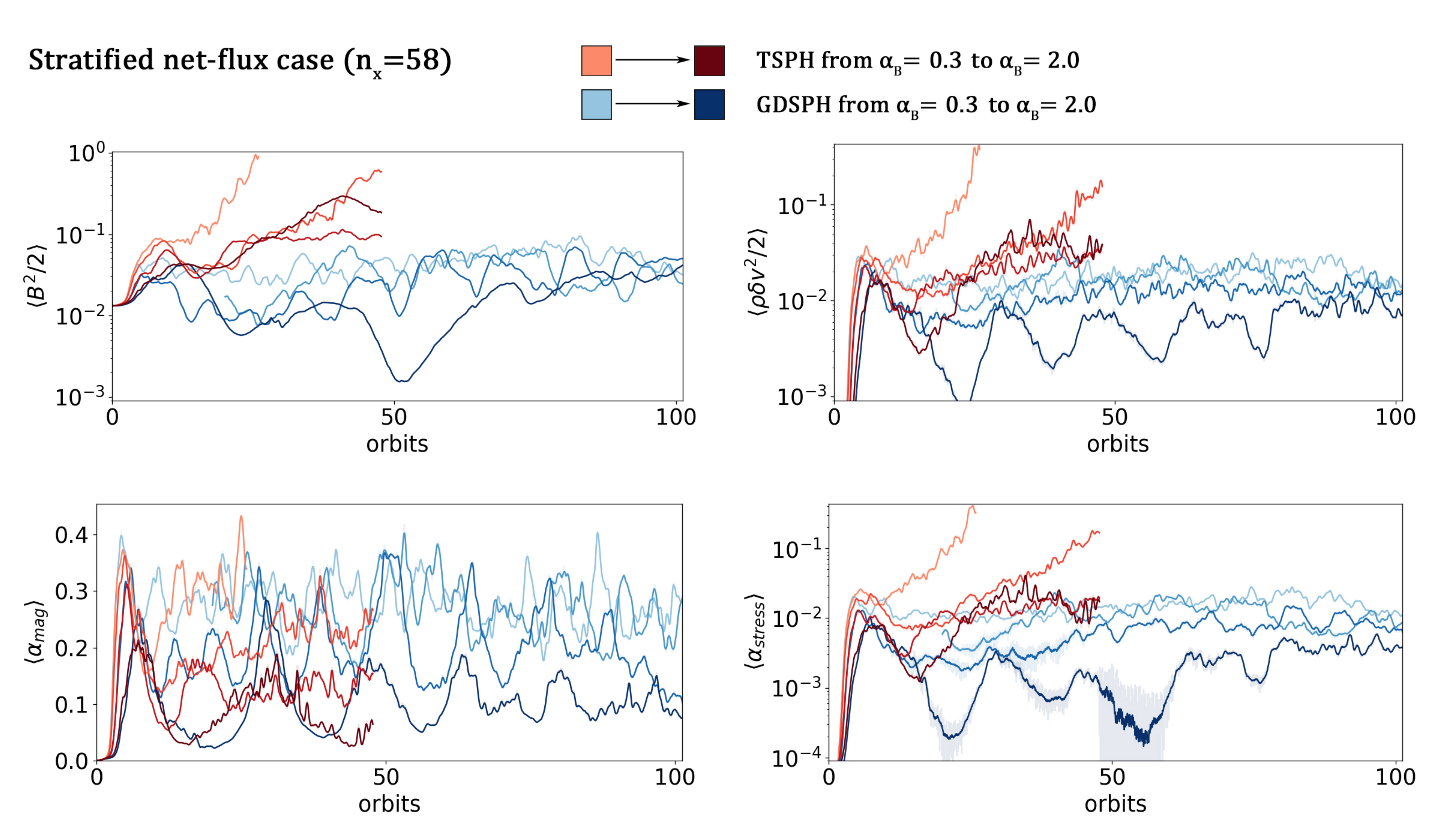}
    \caption{Time evolution of several volume-averaged quantities for the stratified net-flux simulations with varying artificial resistivity($\alpha_B=[0.3,0.5,1.0,2.0]$) at a resolution of $n_x=58$. Magnetic energy(top left), kinetic energy(top right), normalized Maxwell stress(bottom left) and the total stress(bottom right). The red lines show the simulations run with TSPH and blue lines show the runs with GDSPH where the darkness of the line represents the strength of the artificial resistivity. We have smoothed the curves using a Savitzky–Golay filter, the unsmoothed curves can still be seen as very transparent curves.}
    \label{fig:SNFtimeevo}
\end{figure*}

\begin{figure*}[!h]
    \centering
    GDSPH $\alpha_B=0.3$ \qquad \qquad \qquad \qquad \qquad \qquad \qquad \qquad \qquad  TSPH $\alpha_B=0.3$
    \\
    \includegraphics[width=9cm]{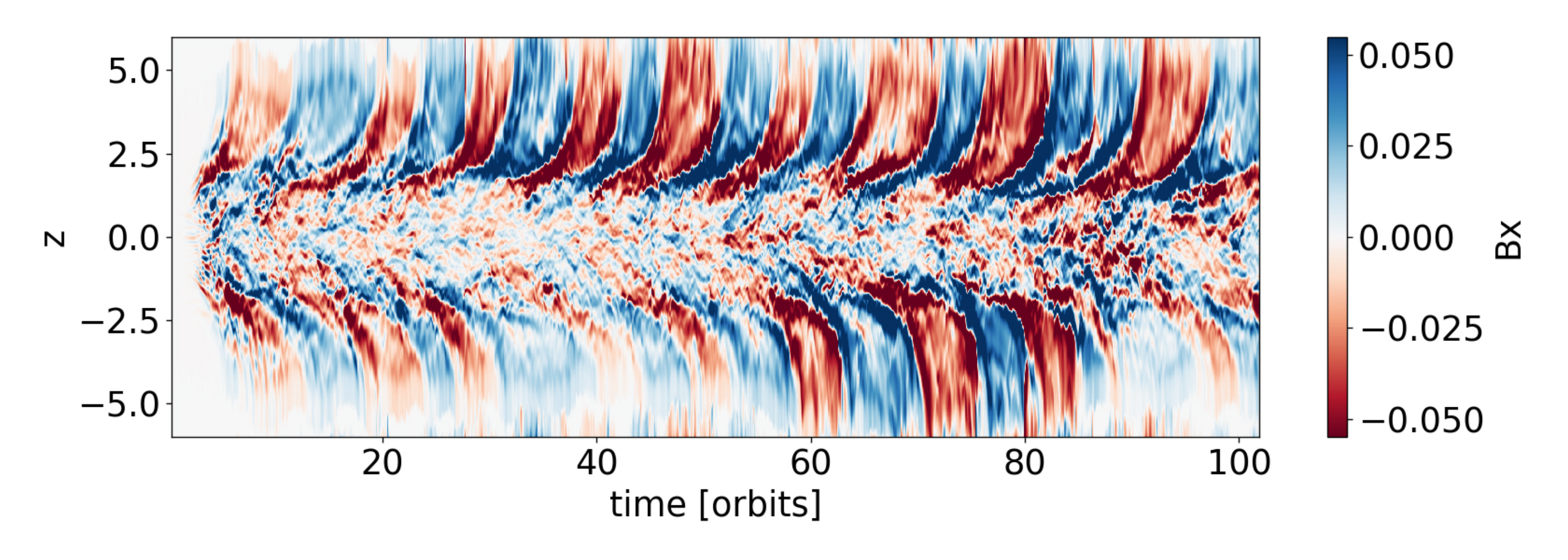}
    \includegraphics[width=9cm]{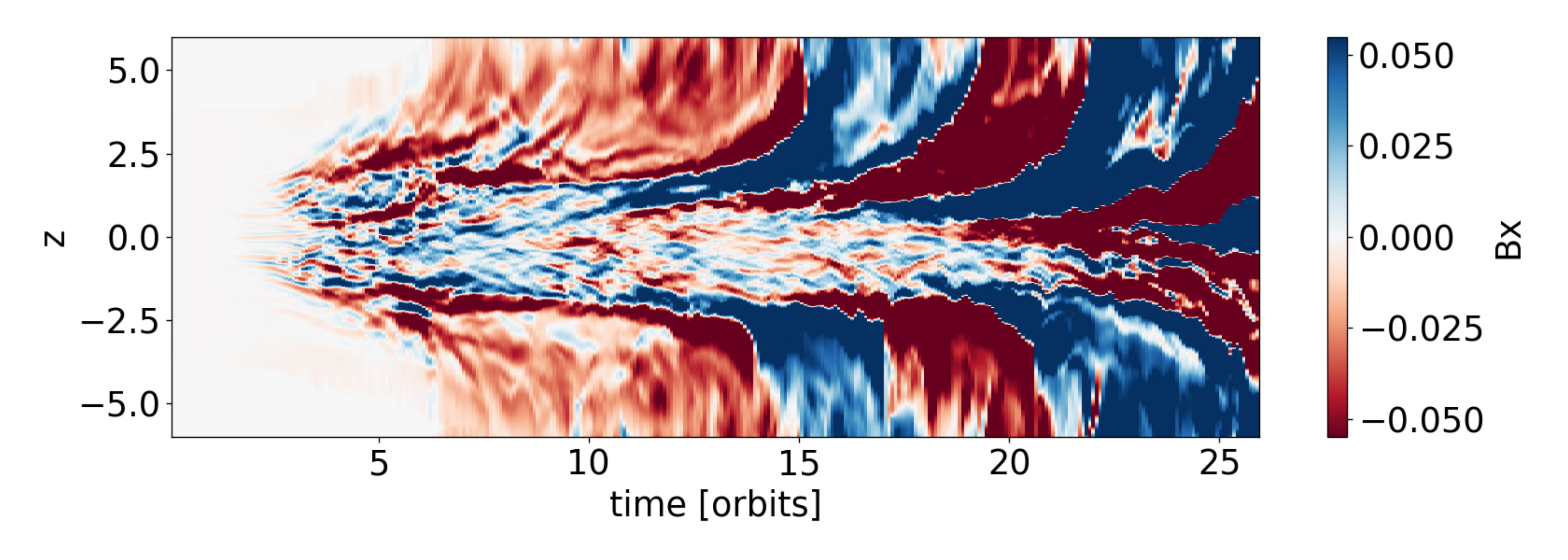}
    \includegraphics[width=9cm]{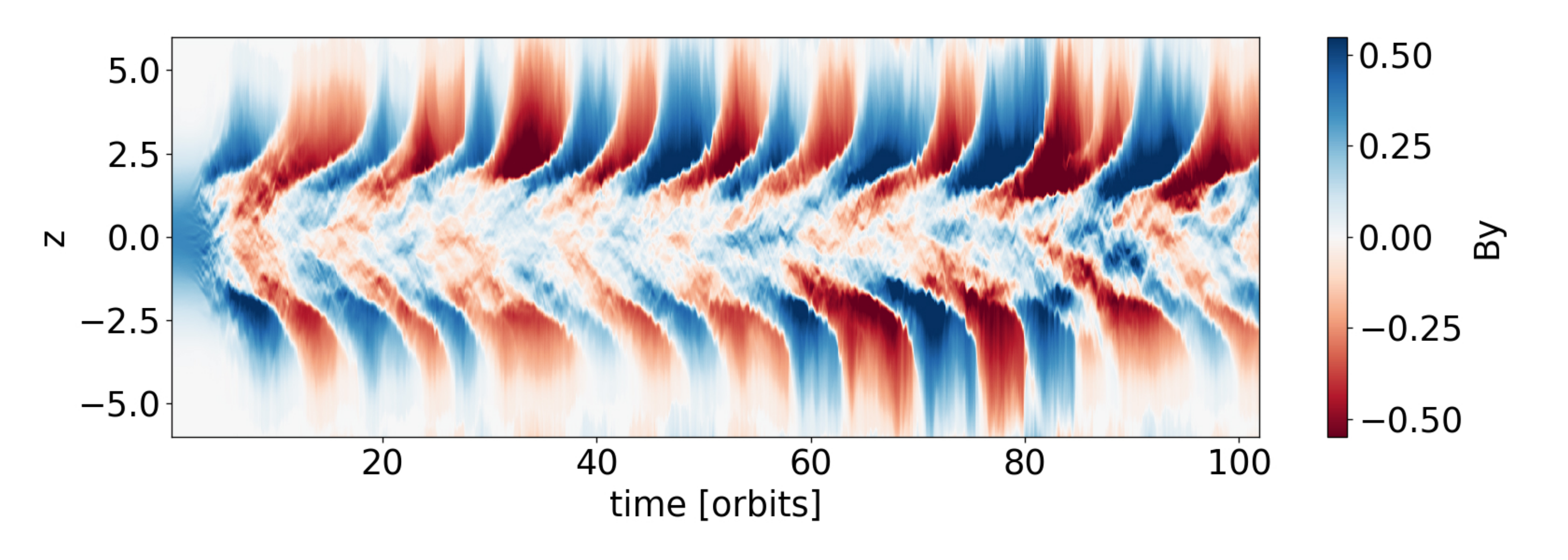}
    \includegraphics[width=9cm]{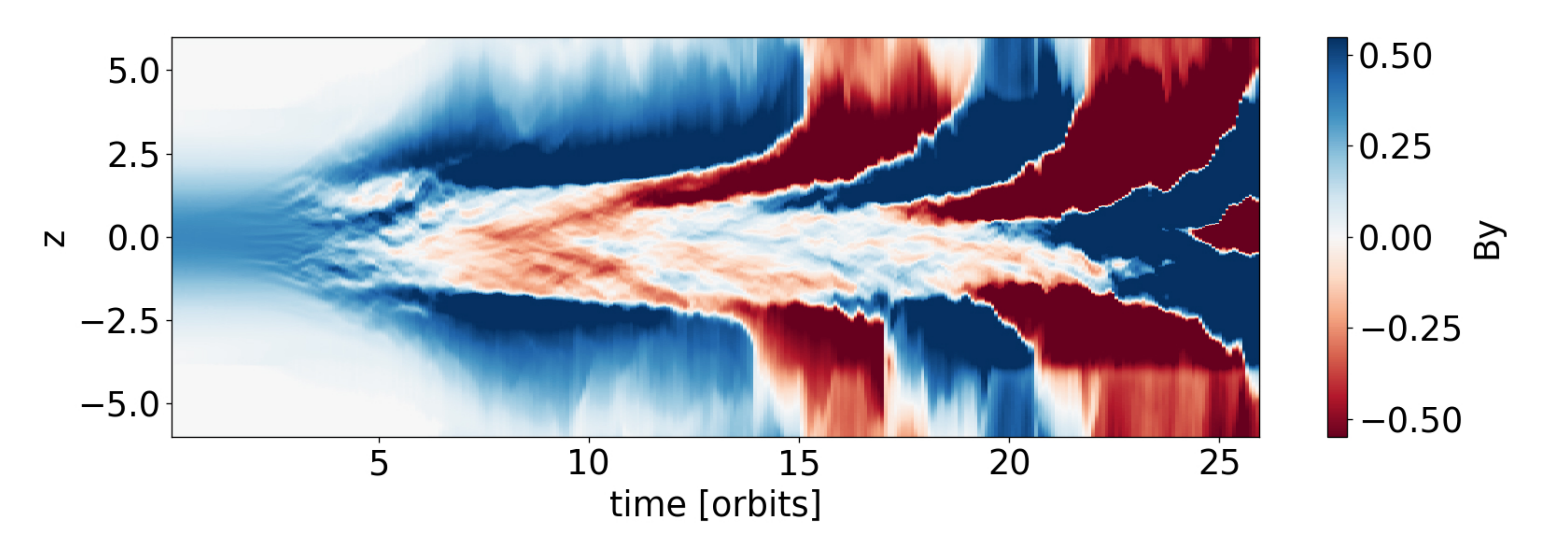}
    \caption{Spacetime diagrams of the stratified net flux simulations, showing the evolution of the horizontal averaged radial(top) and azimuthal(bottom) fields for both GDSPH(left) and TSPH(right) in the case of $\alpha_B=0.3$ with a resolution $n_x=58$. Both GDSPH and TSPH develop the characteristic butterfly diagram. However, the TSPH simulation quickly becomes unstable and exhibits a runaway growth in the magnetic field.}
    \label{fig:SNFrender2}
\end{figure*}
\begin{figure*}[!h]
    \centering
        GDSPH $\alpha_B=1.0$ \qquad \qquad \qquad \qquad \qquad \qquad \qquad \qquad \qquad  TSPH $\alpha_B=1.0$
    \\
    \includegraphics[width=9cm]{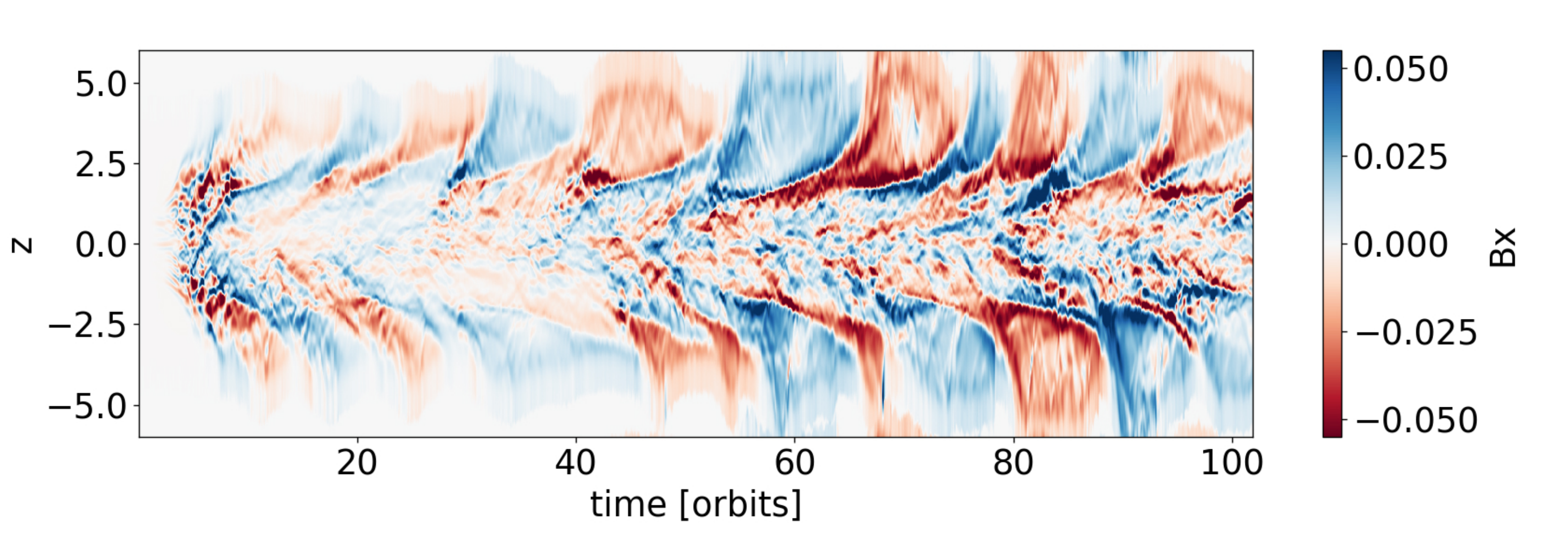}
    \includegraphics[width=9cm]{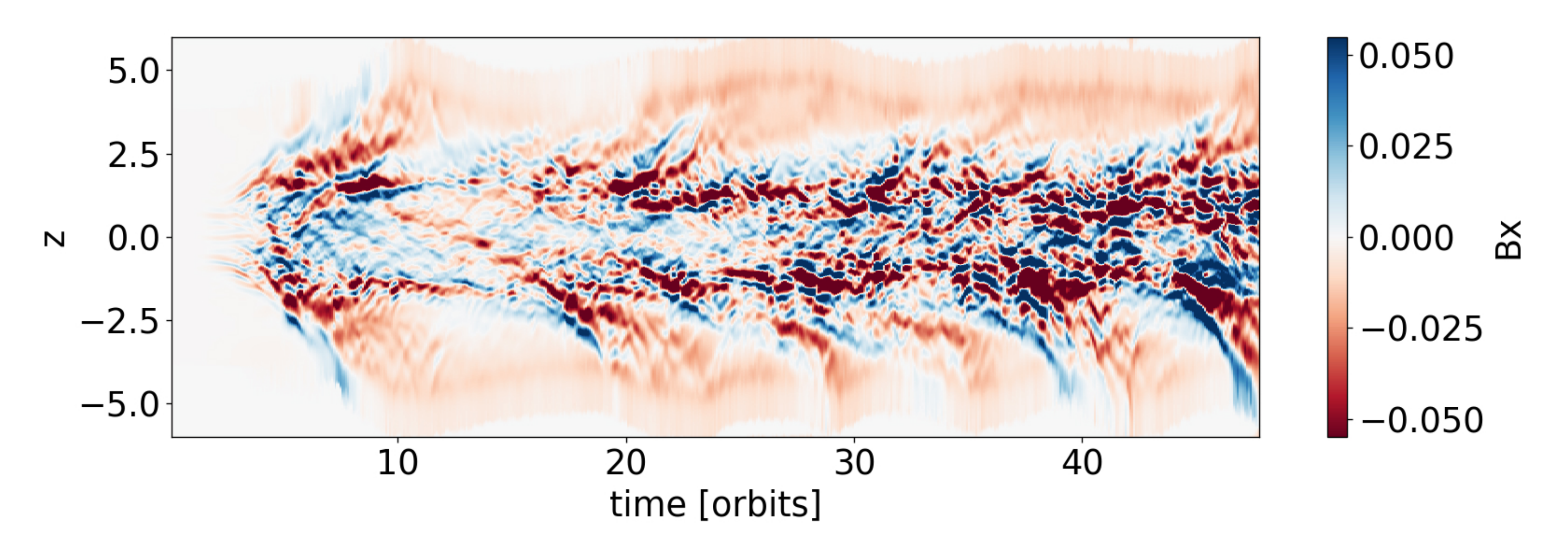}
    \includegraphics[width=9cm]{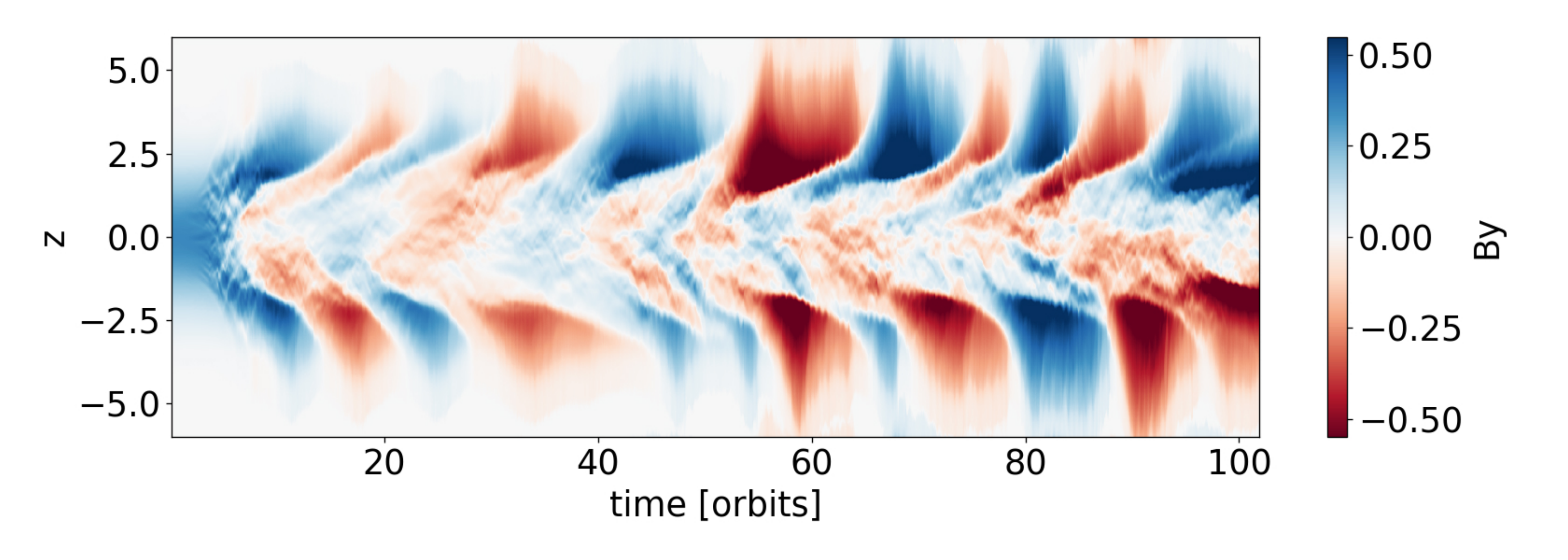}
    \includegraphics[width=9cm]{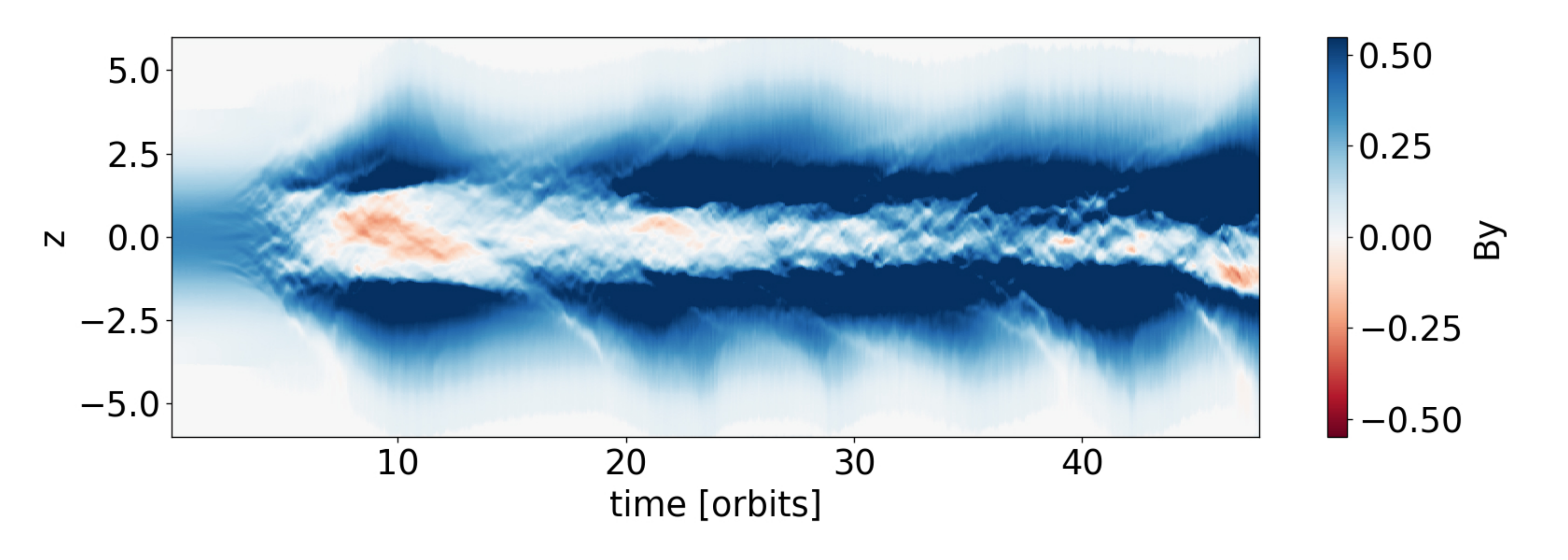}
    \caption{Spacetime diagrams of the stratified net flux simulations, showing the evolution of the horizontal averaged radial(top) and azimuthal(bottom) fields for both GDSPH(left) and TSPH(right) in the case of $\alpha_B=1$ with a resolution $n_x=58$. At this resistivity only GDSPH reproduce the butterfly diagram, where for TSPH a strong positive azimuthal field permeates the disk corona ($\lvert z \rvert>2$). The azimuthal field is additionally amplified as the simulation goes on and starts to propagate into the central disk region.}
        \label{fig:SNFrender3}
\end{figure*}
\begin{figure}[!h]
    
    \centering
    GDSPH \qquad \qquad \qquad \qquad \quad TSPH
    \\
    \includegraphics[width=4cm]{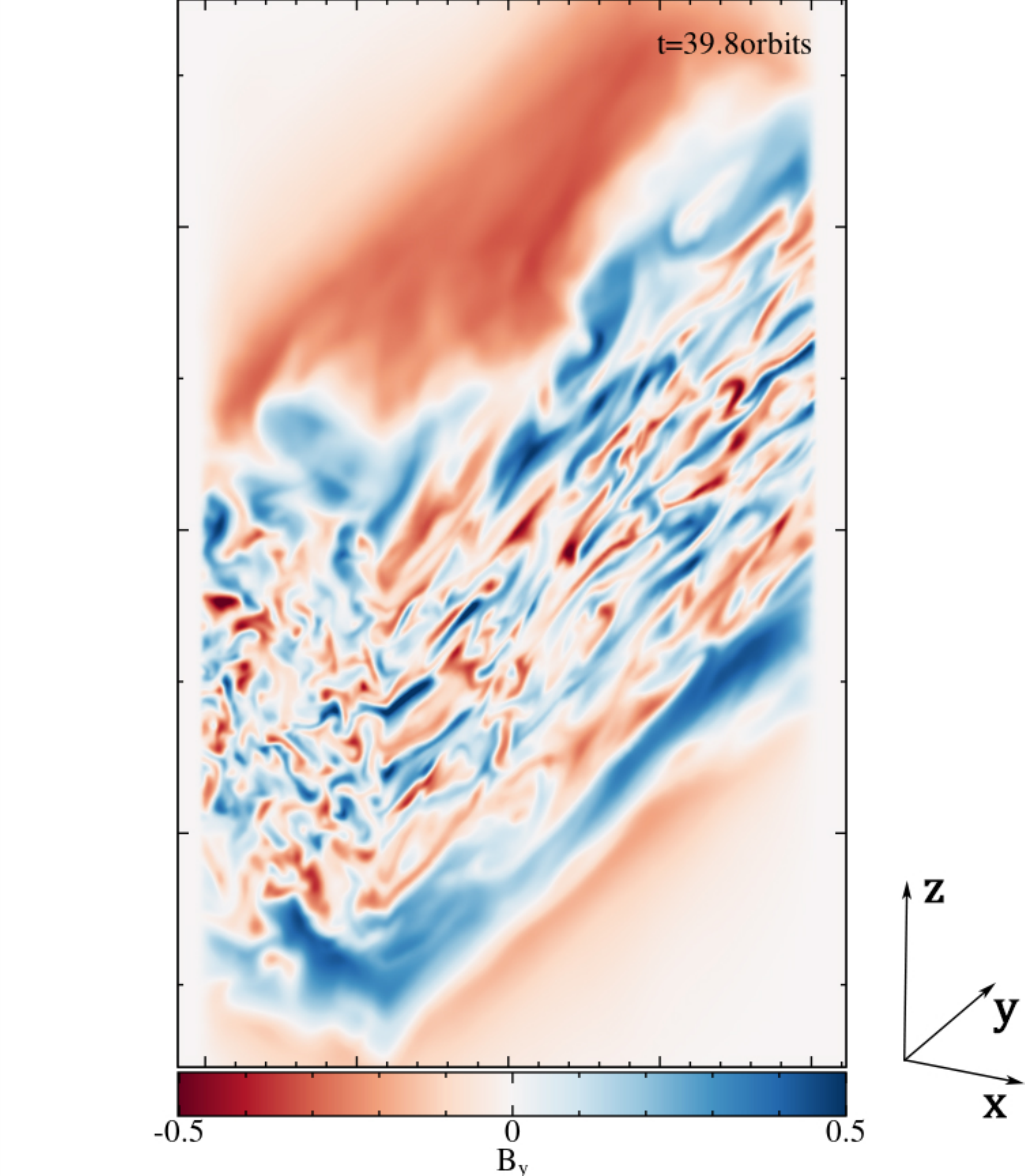}
    \includegraphics[width=4cm]{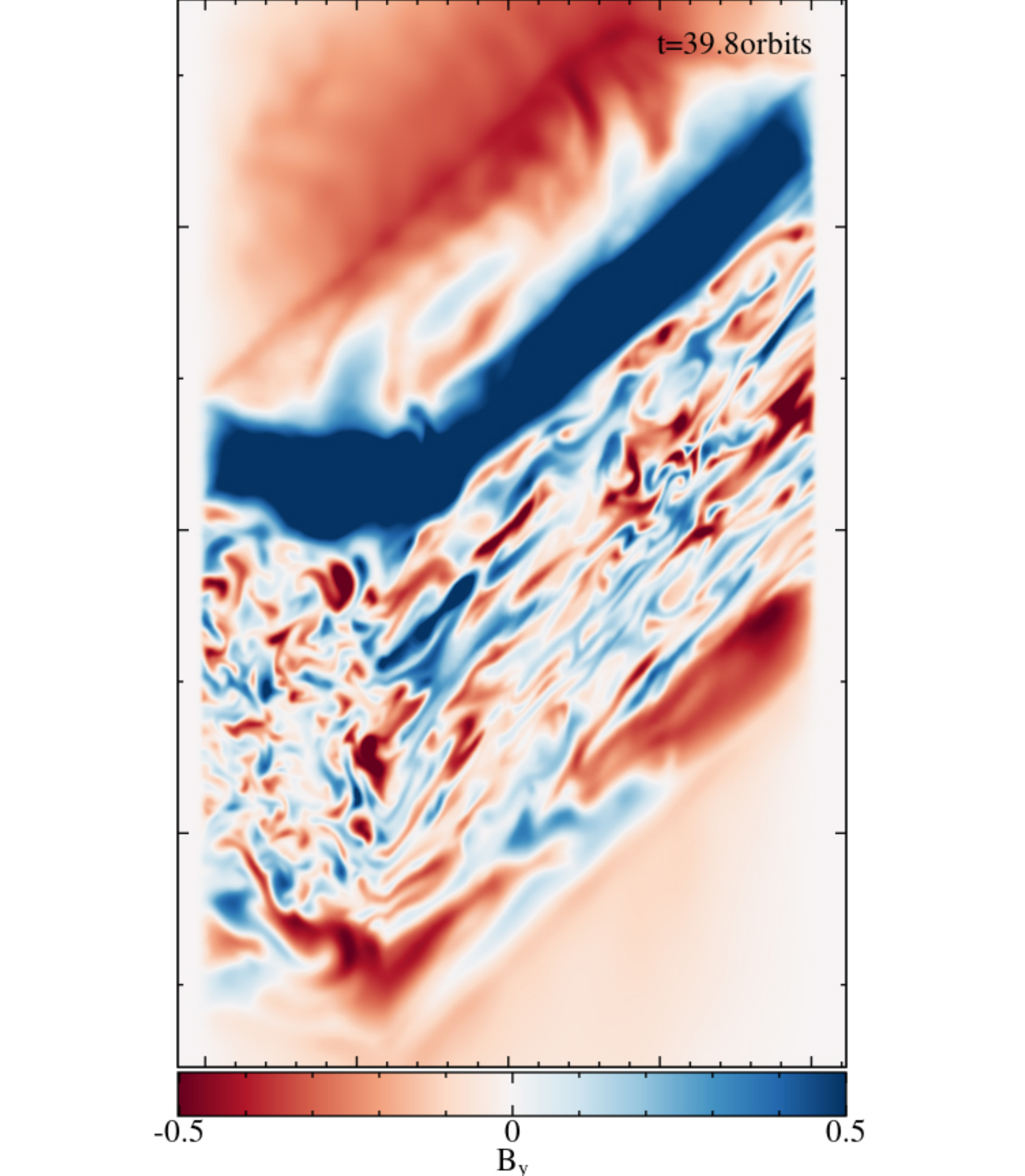}
    \includegraphics[width=4cm]{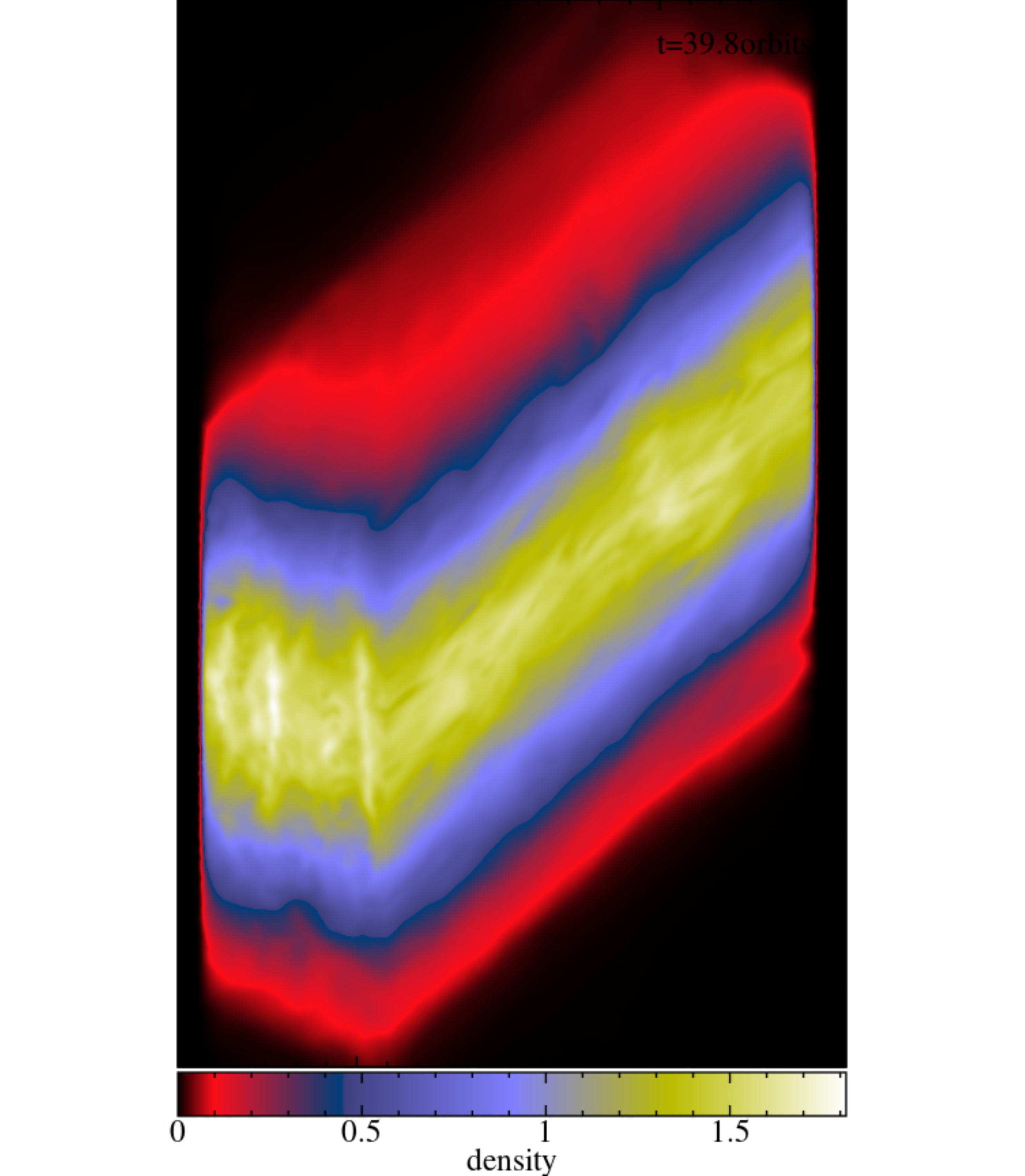}
    \includegraphics[width=4cm]{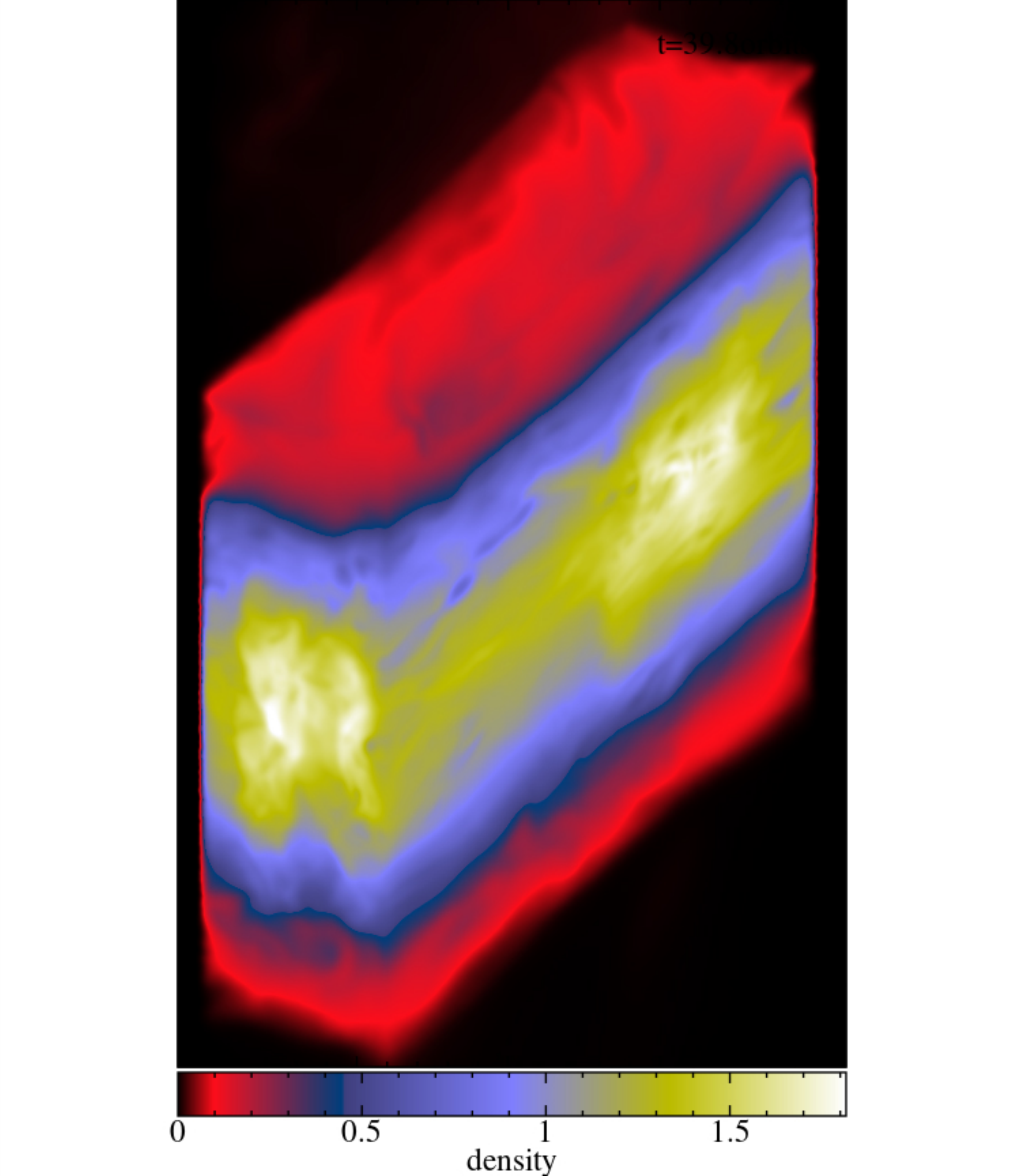}
    \caption{Stratified shearing box simulation with net-flux, shows the surface rendering of the azimuthal field (top) and the density (bottom). Left show the case for GDSPH and right shows the case with TSPH. In the TSPH case we can see that there is excessive growth in the magnetic field in the outer part of the disk.}
    
    \label{fig:SNFrender}
\end{figure}
\begin{figure*}[!h]
    \centering
    \includegraphics[width=\hsize]{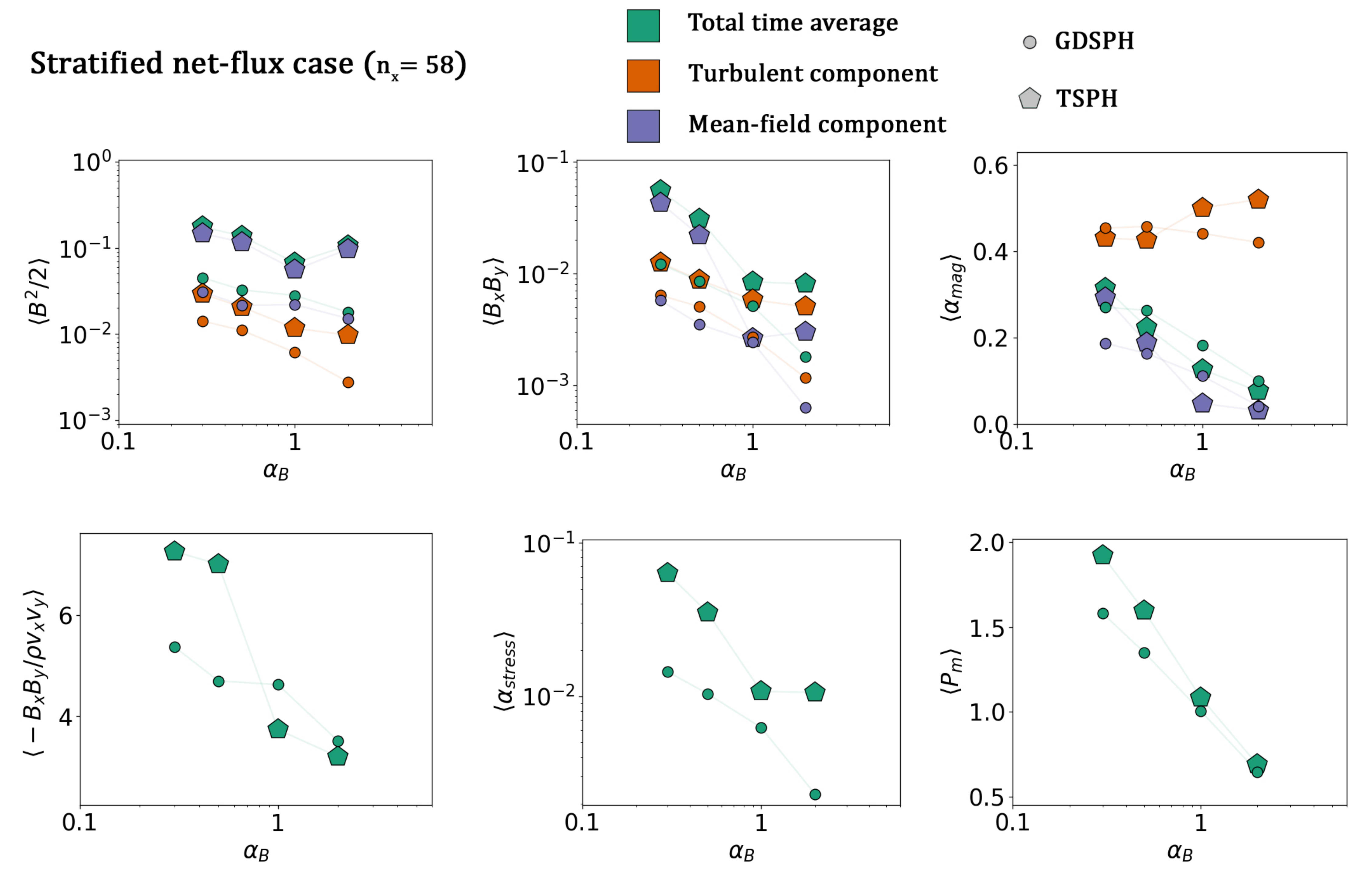}
    \caption{Time-averaged values of several quantities for all our stratified net-flux simulations as a function of the artificial resistivity ($\alpha_B$) at a resolution of $n_x=58$. From the top left to bottom right, we plot the magnetic energy density, the Maxwell stress, the normalized Maxwell stress, the ratio between Reynolds and Maxwell stresses, the total stress, the estimated numerical Prandtl number. For some quantities we have plotted the total time average (shown in green), the time average of the turbulent component(shown in orange) and the time average of the mean component(shown in blue). The circles represent the simulations run with GDSPH while the diamonds represent the simulations run with TSPH.}
    \label{fig:SNFaverage}
\end{figure*}
\begin{figure*}[!h]
    \centering
    \includegraphics[width=16cm]{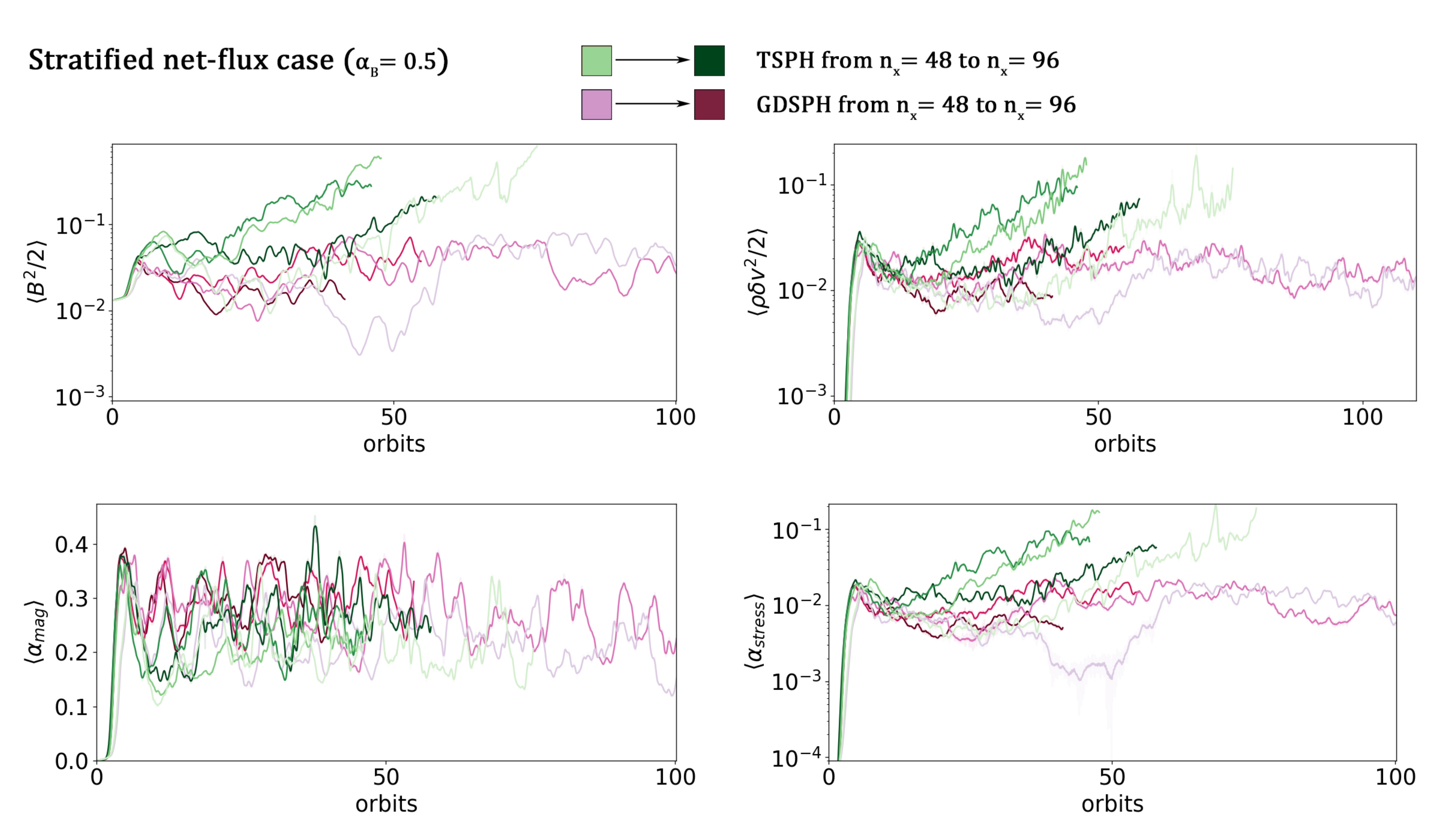}
    \caption{Time evolution of several volume averaged quantities for the stratified net-flux simulations with varying resolution ($n_x=[48,58,74,93]$) and the artificial resistivity coefficient set to $\alpha_B=0.5$. Magnetic energy(top left), kinetic energy(top right), normalized Maxwell stress(bottom left) and the total stress(bottom right). The green lines show the simulations run with TSPH and purple lines show the runs with GDSPH. The darkness of the line indicate the resolution, where the darkest line represents the highest resolution.}
    \label{fig:SNFtimeevores}
\end{figure*}
\begin{figure*}[!h]
    \centering
    \includegraphics[width=15cm]{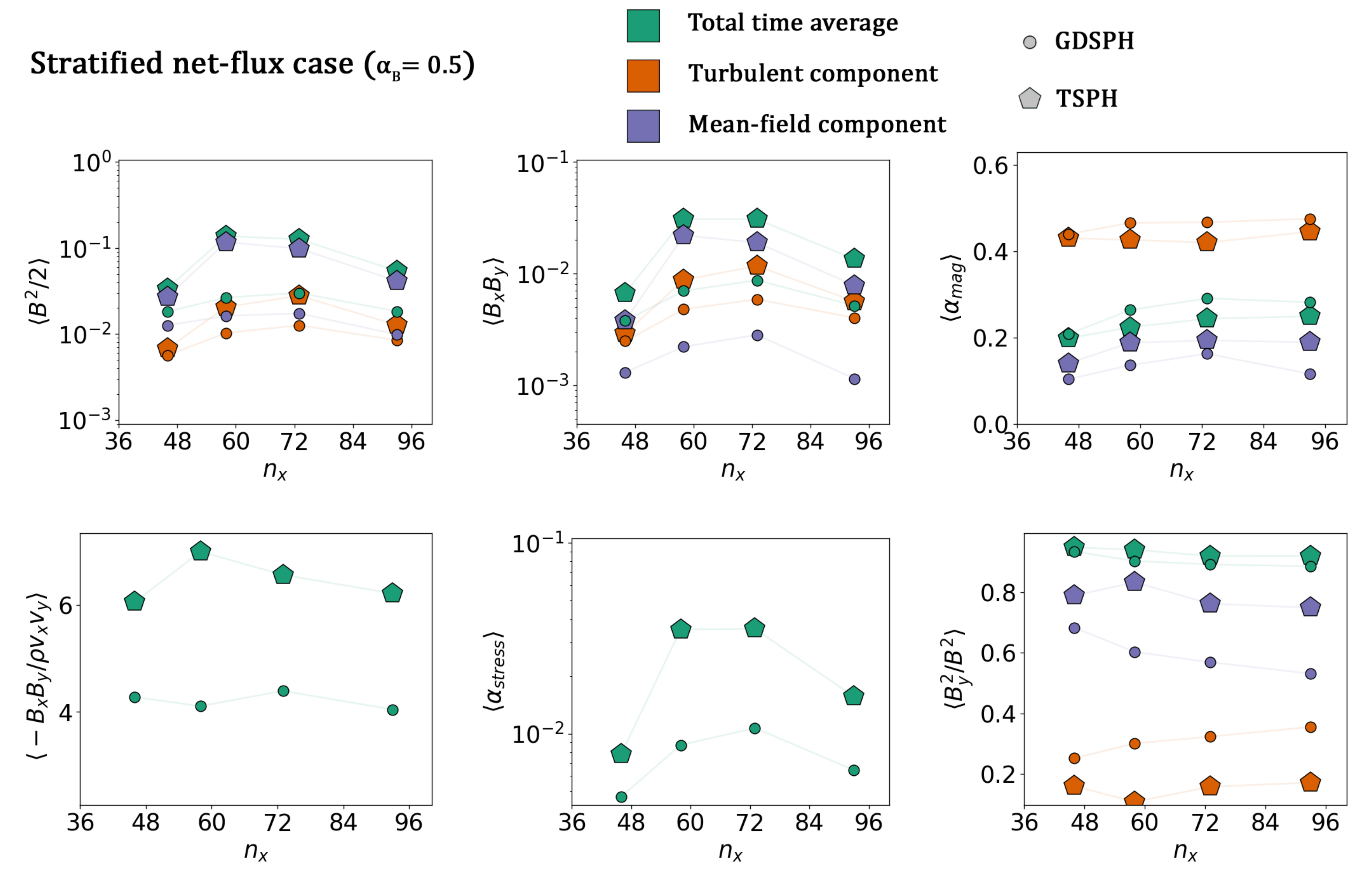}
    \caption{Time-averaged values of several quantities for all our stratified net-flux simulations with artificial resistivity coefficient set to $\alpha_B=0.5$, plotted over the resolution (particles along the x-axis). From the top left to bottom right, we plot the magnetic energy density, the Maxwell stress, the normalized Maxwell stress, the ratio between Reynolds and Maxwell stresses, the total stress, the ratio between azimuthal and total magnetic field energy. For some quantities we have plotted the total time average (shown in green), the time average of the turbulent component(shown in orange) and the time average of the mean component(shown in blue). The circles represent the simulations run with GDSPH while the diamonds represent the simulations run with TSPH.}
    \label{fig:SNFaverageres}
\end{figure*}
\begin{figure*}[!h]
    \centering
    \includegraphics[width=\hsize]{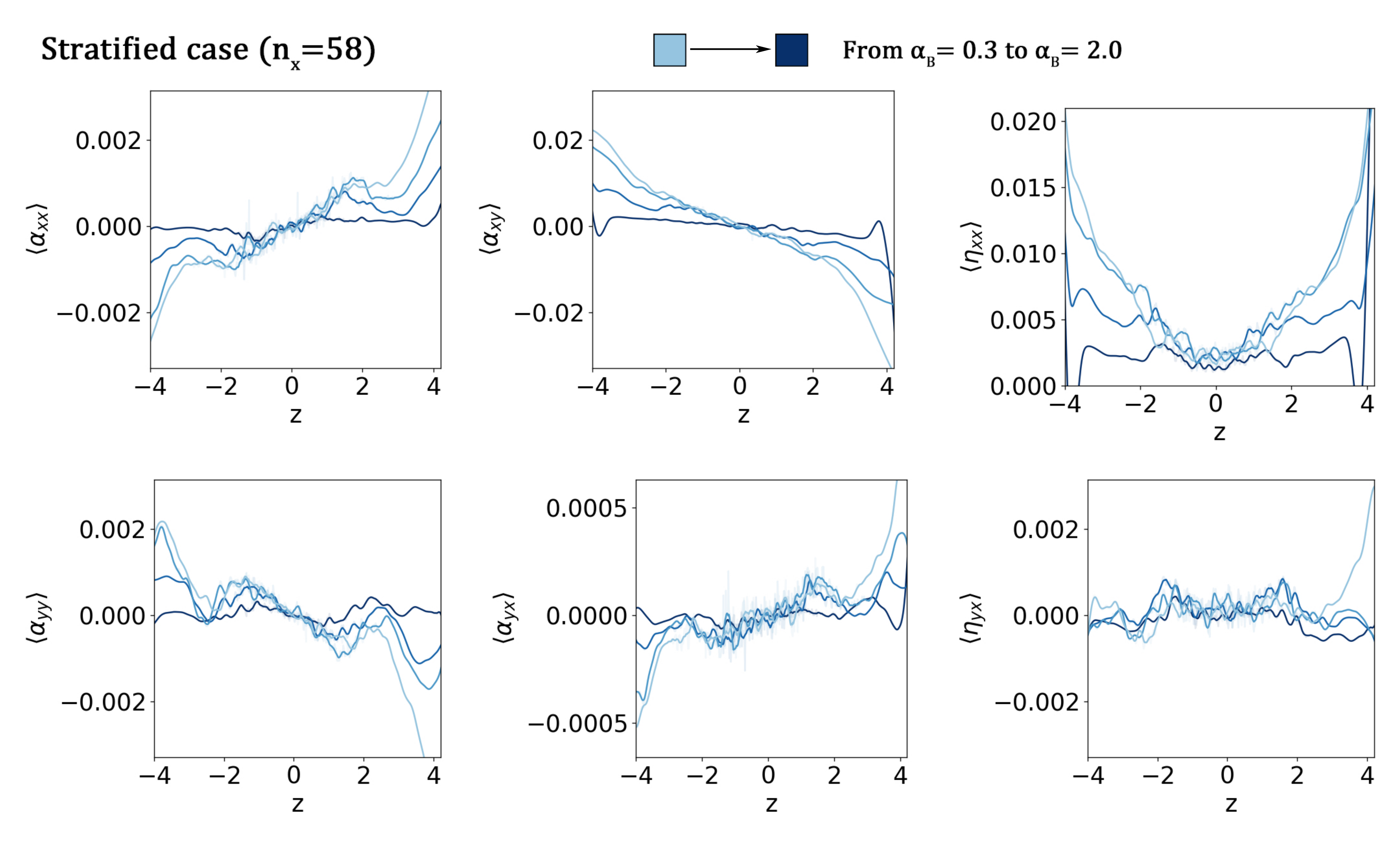}
    \caption{The horizontal time-averaged turbulent transport coefficient in the z direction from the stratified net-flux cases. The darkness of the curves is determined by the strength of the artificial resistivity parameter, $\alpha_B=0.3,0.5,1.0,2.0$. We can see that we have a negative $\alpha_{yy}$ effect that is negative (positive) in the top (bottom) half of the box. This enables the $\alpha\omega$-dynamo to operate efficiently within the central region. $\alpha_{xx}$ on the other hand has a positive gradient and will act against the shear term. Turbulent pumping advects magnetic fields outwards from the central region with velocity $\gamma_z=\frac{1}{2}(\alpha_{yx}-\alpha_{xy})$. However, the $\alpha_{xy}$ magnitudes differ significantly to what is expected, which is likely caused by correlations with $\Bmx$ due to the shear term as explained in section \ref{subsec:dynamo}.
    In the central region we find that at all resistivities we have a positive value for $\eta_{xx}$ and $\eta_{yx}$, which is similar to our result of the unstratified case. In general we find that the behavior of the transport coefficients are similar to previous simulations of the stratified MRI \citep{1995LNP...462..385B,2002GApFD..96..319B,2008AN....329..725B,2016MNRAS.456.2273S,2010MNRAS.405...41G}.
    }
    \label{fig:SNFtransport}
\end{figure*}

The stratified NF simulations represent a more realistic and complex situation than the unstratified case, as it includes the vertical tidal component (final term in \eq\ref{eq:shearapprox}), which results in the following density stratification:
\begin{equation}
\label{eq:stratdensdist}
    \rho=\rho_0 e^{-\frac{z^2}{H^2}}.
\end{equation}{}
Here, H is the scale height and is set by $H=c_s/\Omega=1.0$. As we adopt an isothermal equation of state we do not have to worry about the scale height changing during the simulation. In previous studies, the developed MRI turbulence shows a periodic dynamo cycle, where large-scale magnetic fields emanate from the central region and migrate outwards to the disk corona, growing in strength as they do so. This flips the sign of the field within the central region and the process is repeated. As we mentioned in the introduction, recently it has been shown that SPH develops unphysically strong azimuthal fields in simulations of the stratified shearing box \citep{2019ApJS..241...26D}. In this section, we will further investigate this case with a larger array of different numerical dissipation parameters and resolutions. In addition, we will compare the results from TSPH to the newly developed GDSPH, which has been shown to improve performance in cases involving large density gradients \citep{2020A&A...638A.140W}. We set up the simulation following \cite{2019ApJS..241...26D}, in which they use a shearing box of length $L=(\sqrt{2},4\sqrt{2},24)$ together with an azimuthal magnetic field:
\begin{equation}
    B=\sqrt{\frac{2P}{\beta}}\hat{y}.
\end{equation}
Here, the initial plasma beta is set to $\beta=25$ throughout the box and, as the pressure will vary with density as $P=\rho c_s^2$, we will begin with a magnetic field that varies in the vertical direction.
We run the simulations at four different resolutions $[n_x,N]=[46,1.6 \cdot 10^6], [58,3.1 \cdot 10^6],[73,6.2 \cdot 10^6], [93,12.8 \cdot 10^6]$, where the two lower resolution cases are the same as the ones run in \citep{2019ApJS..241...26D}. As the resolution is adaptive, we will have more resolution in the inner region of the disk (which is where the MRI turbulence is sustained) and less resolution outside in the disk corona. Using \eq~\ref{eq:quality} we can see that we resolve $\lambda_{MRI}$ with an average initial quality parameter of $Q_{mid}=[43,55,68,90]$ in the midplane for each respective resolution. We carry out several simulation at a resolution of $n_x=58$, where we vary the artificial resistivity coefficient $\alpha_B=[0.3,0.5,1.0,2.0]$, where $\alpha_B=0.5$ is the code default value. For all the simulation cases we run one with TSPH and one with GDSPH for comparison. Due to the outflow boundaries, there is mass loss from the simulation, which leads to a flattening of the density profile. This means that resolution will gradually be reduced as time goes on, which is why we at most run our simulation for about 100 orbits. The high-resolution cases are also very computationally costly and are stopped at a somewhat earlier time. The results of the simulations are shown in \fig~\ref{fig:SNFtimeevo} to \ref{fig:SNFtransport}.
\\ \\
In \fig~\ref{fig:SNFtimeevo} we can see the time evolution of the magnetic energy, kinetic energy, normalized Maxwell stress, and the total stress for a resolution with $n_x=58$. From this figure, we can see that all of the TSPH simulations exhibit an unphysical growth in the magnetic energy density, similar to what was seen in \cite{2019ApJS..241...26D}. These simulations are stopped when they reach roughly an average plasma beta value of $\beta=1$, which acts as a confirmation of erroneous growth. On the other hand, all the GDSPH simulations remain stable and those runs with moderate artificial resistivities all reach saturated magnetic energy and stress levels that are similar to what has been seen in the literature \citep{2010ApJ...708.1716S,2011ApJ...730...94S}. To get a closer look at what is going on, we have in \fig~\ref{fig:SNFrender2} plotted the time-space evolution of the horizontal averaged radial and azimuthal fields for both GDSPH and TSPH in the case of $\alpha_B=0.3$ with a resolution $n_x=58$. Both GDSPH and TSPH develop the characteristic butterfly diagram, where the azimuthal fields are buoyantly transported outward and periodically flip signs in the central region. However, while the GDSPH case stably continues this behavior for over 100 orbits, the TSPH case quickly becomes unstable and exhibits a runaway growth. Increasing the resistivity to $\alpha_B=1.0$ does not help stabilize the TSPH scheme, as can be seen in \fig~\ref{fig:SNFrender3}. The butterfly diagram is gone and instead, a strong positive azimuthal field permeates the disk corona ($\lvert z \rvert>2$). The azimuthal field is additionally amplified as the simulation goes on and starts to propagate into the central disk region. The failure of buoyantly ejecting the positive fields in the disk corona is due to the magnetic field growing strong enough to stabilize the region (magnetic tension suppresses the bending of field lines). The time-space diagram of the TSPH case in \fig~\ref{fig:SNFrender3} is reminiscent of the result presented in \cite{2019ApJS..241...26D} where a similar magnetic field growth was observed. The GDSPH case, on the other hand, still exhibits the butterfly diagram at higher resistivity, but with a longer periodic cycle for the flipping of the magnetic field (especially at early times). The TSPH case still has a very active and fluctuating radial field within the central region, and this is also reflected in the the normalized Maxwell stresses in \fig~\ref{fig:SNFtimeevo}, where the values of the TSPH cases remain similar to the GDSPH runs with the same $\alpha_B$. This is further highlighted in \fig~\ref{fig:SNFrender}, where we can see a rendering of the magnetic field and density within the box for both the TSPH and GDSPH cases. Both simulations exhibit a very similar central region, while the TSPH have significantly stronger azimuthal fields in the outskirts. As TSPH and GDSPH mainly differ at density gradients, it makes sense that the issue of the unphysical growth seems to lie in the outer region of the disk (beyond $\lvert z \rvert>1$) where we have lower resolution and a significant density gradient.
\\ \\
\fig~\ref{fig:SNFaverage} shows time averages of different quantities. In general, as we decrease the resistivity, the magnetic energy and stress increases for both the TSPH and GDSPH runs. The total stress reaches a time-averaged value of around $\alpha_{stress} \approx 10^{-2}$ and a normalized Maxwell stress of around $\alpha_{mag}=0.3$ for the low resistivity cases, similar to previous result from the literature \citep{2010ApJ...708.1716S,2011ApJ...730...94S}. Looking at the mean and turbulent components of the magnetic energy density, we can see that the mean-field component is the dominant part, but with an increasing fraction from the turbulent field as we decrease the resistivity. The TSPH runs develop a much higher magnetic energy density than the GDSPH cases, with a highly dominating mean-field component, which comes mainly from the strong azimuthal fields in the corona. For the Maxwell stress, in our GDSPH simulations the turbulent and mean-field components contribute a similar amount to the total Maxwell stress. The normalized Maxwell stresses have similar values for both GDSPH and TSPH, with the turbulent component of around $0.45$ and largely independent of the resistivity, but the mean-field component increases  with decreasing resistivity.
\\ \\
From \fig~\ref{fig:SNFaverage} we can also see that the ratio between the Maxwell stress and Reynolds stress show a value of around $5.0$ for the GDSPH cases with an increasing trend for lower resistivity. For the TSPH runs, this ratio shoots up for the low resistivity cases to a value of around 8. The numerical Prandtl number has a steady increase as we decrease the resistivity. For GDSPH it goes from a value of around $0.7$ for $\alpha_B=2.0$ to a value of $1.6$ for $\alpha_B=0.25$. In TSPH the numerical Prandtl number is larger but not by much. The average divergence error for both GDSPH and TSPH remains close or below a value of $\epsilon_{div,err}\approx 10^{-2}$.
\\ \\
To investigate the effect of resolution in the stratified simulations, we perform a resolution study with the following radial resolutions $n_x=[48,58,73,93]$. In these simulations, we use the code default artificial resistivity coefficient of $\alpha_B=0.5$. The high-resolution cases ($n_x=73,93$) are very costly, so we have only run them for about $60$ orbits. In \fig~\ref{fig:SNFtimeevores} we can see the time evolution of the magnetic energy, kinetic energy, normalized Maxwell stress, and the total stress for TSPH and GDSPH at different resolutions. From this figure, we can see that all the TSPH cases eventually become unstable, and the lowest resolution cases "survives" the longest. All of the GDSPH cases show a stable behavior, with only the lowest resolution case having a period of low stress before increasing to similar levels as the higher resolution cases. A significant early difference that can be seen between TSPH and GDSPH is shown in the magnetic energy density, where all the TSPH have a much larger initial "bump" than the GDSPH curves which flatten out after the initial increase. This larger initial bump correlates to stronger magnetic fields near the density contrast at around $\lvert z \rvert \approx 1$. 
\\ \\
In \fig~\ref{fig:SNFaverageres} we can see that the values of magnetic energy density and stress remain fairly flat, with a slight initial increase with resolution but then a slight decrease for our highest resolution. It is not clear to why we see a decrease in the stress for the highest resolution. It could simply be a stochastic phenomena that would flatten out if we ran it for longer. However, both GDSPH and TSPH follow a similar curve, pointing towards a real effect. We can see that the turbulent component does become a more dominant part in both the energy density and in the stress as we increase resolution. The normalized Maxwell stress has a slight increase with resolution, going from around $\alpha_{mag}=0.2$ to $\alpha_{mag}=0.3$ and the total stress for GDSPH lies between $\alpha_{stress}=10^{-3}\leftrightarrow10^{-2}$ which is in accordance to the values presented by \cite{1996ApJ...463..656S} and \cite{2010ApJ...708.1716S}. We can see that the ratio between the Maxwell stress and Reynolds stress is almost independent of resolution giving a value of $5.0$ for the GDSPH cases and around $7.0$ for TSPH. Similar to the unstratified case, we can in \fig~\ref{fig:Pmresdepend} see that we have a slow but linear increase in the numerical Prandtl number with resolution, going from around $P_m=1.3$ to $P_m=1.5$ for the GDSPH cases. The average divergence error slightly decreases with resolution for both TSPH and GDSPH with values around $\epsilon_{div,err}\approx 10^{-2}$.
\\ \\
In \fig~\ref{fig:SNFtransport} we show the horizontal time-averaged turbulent coefficients as a function of z. As we mentioned in the introduction, the dynamo action within stratified disks includes several mechanisms that can act in different sections of the disk. We find that the $\alpha_{xx}$, $\alpha_{yy}$, $\alpha_{xy}$ and $\alpha_{yx}$ all have a continuous gradient, with anti-symmetric behavior around the mid-plane. The behavior of $\alpha_{yy}$ determines the effectiveness of the $\alpha\omega$-dynamo, as it operates on the radial mean fields (see \eq \ref{eq:dBdtx}). We can see that $\alpha_{yy}$ has a negative gradient with negative values above the midplane, this is consistent with \cite{1995LNP...462..385B,2002GApFD..96..319B,2008AN....329..725B,2016MNRAS.456.2273S}, and is indicative of an effective alpha effect for turbulent shear flows. The $\alpha_{xx}$ has a positive gradient and will act against the shear term, however, the shear term is the dominant part of the induction equation for the toroidal mean field. The off diagonal terms of $\alpha$ is often interpreted as a turbulent/diamagnetic pumping term $\gamma_z=\frac{1}{2}(\alpha_{yx}-\alpha_{xy})$. For our case, $\gamma_z$ is positive above the mid-plane and negative below, meaning that we have a net transport of mean-fields away from the mid-plane. The $\alpha_{xy}$ and $\alpha_{yx}$ coefficients have the opposite sign, which is similar to the results from \cite{2016MNRAS.456.2273S}, but different from other works by \cite{2008AN....329..725B,2010MNRAS.405...41G}, where the two have the same sign. However, $\alpha_{xy}$ and $\alpha_{yx}$ usually have similar magnitude in earlier works which is not observed in our cases. The $\eta_{xx}$ coefficient and $\eta_{yx}$ are generally positive, which is similar to what was found by \cite{2008AN....329..725B,2010MNRAS.405...41G} but contrary to \cite{2016MNRAS.456.2273S}.

\section{Discussion}
\label{sec:discussion}
In this paper, we performed a plethora of shearing-box MRI simulations using SPH. For the unstratified NF case, we reproduced the results from previous studies in the literature, albeit with slightly larger $\alpha_{mag}$ values and in general larger mean-field stresses. We attribute this primarily to the use of a smaller box, which increases the amplitude and frequency of channel modes and causes large bursts of magnetic energy and stress levels.
\\ \\
We demonstrate that the saturation of the unstratified ZNF simulations is highly dependent on the numerical Prandtl number. Our simulations with a Prandtl number above $2.5$ achieve saturation for at least 200 orbits. To further illustrate that the MRI saturation depends on the numerical Prandtl number, rather than simply on the resistivity, we ran a simulation with a forced Prandtl number of $P_{m,AV}=0.25$ but with a low artificial resistivity coeffecient $\alpha_B=0.25$. This simulation does not saturate even though having a very low numerical resistivity. This confirms that the dependencies on the Prandtl numbers found in \cite{2007A&A...476.1113F} still holds true for numerical Prandtl numbers within SPH. The saturation levels in energies and stresses are also mainly dependent on the Prandtl number. However, for saturated simulations with the same Prandtl number, the magnetic energy and stress are slightly higher in cases with varying artificial resistivity than the ones with varying artificial viscosity. This is also shown in \fig~\ref{fig:UZNFtimeevo} and \ref{fig:UZNFTtimeevo}: when comparing the $P_{m,AR} = 3$ case with $P_{m,AV} = 3$ run, the latter exhibit larger oscillations and lower saturation levels.  
\\ \\
We do not observe a decrease in stress with increasing resolution as found in previous studies with Eulerian codes. Although the stresses are highly dependent on numerical Prandtl number, they have a weak dependency on the resolution at a fixed Prandtl number, either increasing or staying roughly at the same stress level with increasing resolution. A possible explanation is that the numerical Prandtl number in Eulerian codes is not independent of resolution. This is contrary to studies by \cite{2007A&A...476.1123F} and \cite{2009ApJ...690..974S}, where the authors found the numerical Prandtl number to be almost independent of resolution ($P_m\approx1.6$). These studies utilized Fourier transfer functions to compute the energy transfer between different scales. They found that an active MRI can exist in their simulations even though the numerical Prandtl number is lower than the critical value determined in studies with physical dissipation ($P_{m} \sim 2$). Thus, it was concluded that numerical dissipation acts differently to physical dissipation. In this paper, we however find that active turbulence still requires similar critical Prandtl numbers found in studies using physical dissipation. It is likely that the numerical dissipation in SPH is more closely related to physical dissipation than in grid codes, as SPH does not suffer from advection errors. This difference seems to be the reason to why grid codes see a significant reduction in stress with resolution. 
\\ \\
Ideally, one would like the numerical Prandtl number to remain independent of resolution, as this would ensure the correct dynamo behavior if one can resolve the turbulent medium. The Prandtl numbers in our simulations increase with resolution for a fixed artificial resistivity coefficient ($\alpha_{B} = 0.5$) ranging from $\langle P_{m} \rangle \sim 1.5$ for $n_{x} = 48$ to $\langle P_{m} \rangle \sim 2.1$ at $n_{x} = 96$. Although not independent, an increase of $P_{m}$ with resolution ensures that it increases along with the other fluid parameters (e.g., Reynolds number and magnetic Reynolds number). This means that we will have convergent results when modeling most astrophysical fluids. A worse result would have been if the Prandtl number decreased with resolution which could result in a lowering of stress with resolution and finally in the decay of the MRI. In addition, as mentioned in the introduction, the Prandtl number plays a major role in several dynamo mechanisms and is crucial for the saturation of the small-scale dynamo, which for example can be important for  correctly simulating the growth and saturation of magnetic fields within galaxies. Our results highlight the importance of studying the numerical Prandtl number for all numerical schemes beyond MRI, for instance in turbulent boxes at different Mach numbers.
\\ \\
The main difference in our tall boxes compared to \cite{2016MNRAS.456.2273S} is the lack of significant mean-fields, which leads to stress levels that are only slightly larger than the ones in the standard box cases, but much smaller than $\alpha_{stress}=10^{-1}$ seen in \cite{2016MNRAS.456.2273S}. Our simulations do develop similar large-scale patches in the toroidal field, but they are significantly weaker. The small-scale turbulent components are consistent with the \cite{2016MNRAS.456.2273S} results. 
The lack of mean-fields is likely due to the difference seen in $\eta_{yx}$, which was consistently negative for \cite{2016MNRAS.456.2273S} but for all our simulations are either zero or positive. This would effectively lead to less coherent mean-field growth within the box. However, the positive value of $\eta_{yx}$ is consistent with previous studies with the quasi-kinematic approach \citep{1995ApJ...446..741B,2008AN....329..725B,2010MNRAS.405...41G}. For future work, it is worth exploring higher resolutions and different aspect ratios for SPH, to see if higher mean-field growth can be observed.
\\ \\
We demonstrated, for the first time, that the new GDSPH can successfully sustain the turbulence in the stratified shearing boxes for at least 100 orbits without decaying, similar to the simulations using the {\sc Gizmo} code with the meshless finite-mass (MFM) method \citep{2019ApJS..241...26D}. However, the TSPH runs remain unstable for all cases, which confirms the result from \cite{2019ApJS..241...26D}. We conclude that this is partly due to a magnetic flux error, as the energy is continuously increasing together with the magnetic field getting either more and more positive or negative. Global mean-field such as this can either be generated by the outflow boundaries or the monopole currents. Since outflow boundaries tend to expel flux roughly equally, the error is more likely to be caused by monopole currents, together with the gradient errors of TSPH near the density gradient. Compared to GDSPH, TSPH always develops very strong azimuthal fields beyond $\lvert z \rvert >H$. These fields eventually become so strong that they can no longer be buoyantly transported outward, because the critical wavelength is larger than the radial size of box. The negative radial fields in the outskirts will continue to increase the azimuthal field and subsequently the total magnetic energy. In \cite{2019ApJS..241...26D} this unphysical increase of the azimuthal field was partly attributed to the divergence cleaning. However, we argue that this is unlikely the case, because the hyperbolic divergence cleaning is conservative in both energy and volumetric flux, which means that the spreading of the magnetic field due to the cleaning is always symmetrical and no global mean-field can be generated within the box.
\\ \\
In addition, in our simulations the MRI turbulence is sustained longer than the MFM simulation presented in \cite{2019ApJS..241...26D}, in which the turbulence decays around 40 orbits in their fiducial run. Our code has the ability to sustain long-term MRI turbulence similar to the Eulerian codes in previous MRI studies \citep{2010ApJ...708.1716S,2010ApJ...713...52D}. However, as the simulations were terminated earlier in \cite{2019ApJS..241...26D}, it is unclear if the MRI has actually fully died down in that work. We do also observe that the MRI can temporarily dip down to similar values before eventually being re-energized. 
\\ \\ 
We also performed an analysis of the turbulent coefficients for all the simulations presented in this paper. We showed that no $\alpha$-effect was present for the unstratified case as expected. For the stratified case, we have a negative $\alpha_{yy}$ effect that is negative (positive) in the top (bottom) half of the box, which indicates an effective $\alpha \omega$-dynamo. This is similar to what was found in \cite{2008AN....329..725B} and \cite{2016MNRAS.456.2273S}. We find a turbulent pumping that transports the mean-fields away from the central region. However, we note that there is a significant uncertainty in the calculated $\alpha_{yx}$ coefficients due to correlations with the shear term, as explained in section \ref{subsec:dynamo}. The turbulent resistivity $\eta_{xx}$ and $\eta_{yx}$ are found to be positive in all of our simulations, and are consistent with previous quasi-kinematic studies employing the test-field method \citep{1995LNP...462..385B,2002GApFD..96..319B,2008AN....329..725B}.
\\ \\
Using the constrained hyperbolic divergence cleaning scheme with variable cleaning speed from \citet{2016JCoPh.322..326T}, we can keep the divergence error low in all cases. The mean normalized divergence error, $\langle\epsilon_{divB}\rangle = \langle h|\nabla\cdot \Bv|/|B|\rangle$, is typically of order $10^{-2}$.
\\ \\
In conclusion, we find that  
\begin{itemize}
\item SPH can effectively develop the MRI and reproduce many of the values and dependencies seen in previous studies with grid-based codes.  
\item The geometric density SPH (GDSPH) successfully develops the characteristic "butterfly" diagram of the stratified MRI, showing saturated turbulence for at least 100 orbits. The results are similar to MRI simulations with the MFM method, and the turbulence is sustained longer.  
\item The numerical dissipation in SPH is found to act in a similar fashion to physical dissipation. We find a critical Prandtl number of around $P_m \approx 2.5$, which is similar to what grid codes find with physical dissipation. 
\item The saturated stress for a certain numerical Prandtl number is found to be nearly independent of resolution, which is contrary to grid codes where stress is reduced with increased resolution. The results highlight the importance in determining the general behavior of the numerical Prandtl number in different turbulent flows, to ensure a more accurate saturation of the magnetic field. 
\item A major difference can also be seen in the tall, unstratitified, zero net-flux case, where the mean-fíelds are much weaker than a previous study. From the mean-field analysis, we speculate that this might be due to a lack of shear-current effect our simulations. Nevertheless, we find that our transport coefficients are consistent with many previous studies that also do not find an effective shear-current effect. 
\end{itemize}

\bibliographystyle{aa}
\bibliography{references}

\end{document}